\documentclass[useAMS,usenatbib]{mn2e}
\usepackage{graphicx}

\newcommand{\msun}{\mbox{${\rm M}_{\odot}$}}

\title[GCE in SAMs]
{Galactic chemical evolution in hierarchical formation models - I. Early-type
galaxies in the local Universe} 
\author[M. Arrigoni et al.]
{Mat\'{i}as Arrigoni$^1$\thanks{email: arrigoni@astro.rug.nl}, Scott C. 
  Trager$^1$, Rachel S. Somerville$^{2,3}$ and Brad K. Gibson$^4$ \\ 
$^1$Kapteyn Astronomical Institute, University of Groningen, Postbus 800, 
  NL-9700 AV Groningen, The Netherlands\\ 
$^2$Space Telescope Science Institute, 3700 San Martin Drive, Baltimore, 
  MD 21218, USA\\ 
$^3$ Department of Physics and Astronomy, Johns Hopkins University, Baltimore, 
  MD 21218, USA\\
$^4$University of Central Lancashire, Jeremiah Horrocks Institute for 
  Astrophysics \& Supercomputing, Preston, PR1 2HE, United Kingdom} 
\begin{document}

\date{Accepted . Received .}

\pagerange{\pageref{firstpage}--\pageref{lastpage}} \pubyear{2009}

\maketitle

\label{firstpage}

\begin{abstract}
We study the metallicities and abundance ratios of early-type galaxies in cosmological 
semi-analytic models (SAMs) within the hierarchical galaxy formation paradigm. To 
achieve this we implemented a detailed galactic chemical evolution (GCE) model and can 
now predict abundances of individual elements for the galaxies in the semi-analytic 
simulations. This is the first time a SAM with feedback from Active Galactic Nuclei 
(AGN) has included a chemical evolution prescription that relaxes the instantaneous 
recycling approximation. We find that the new models are able to reproduce the observed 
mass-metallicity ($M_{\star}$--[Z/H]) relation and, for the first time in a SAM, we 
reproduce the observed positive slope of the mass-abundance ratio 
($M_{\star}$--[$\alpha$/Fe]) relation.  Our results indicate that in order to 
simultaneously match these observations of early-type galaxies, the use of both a very 
mildly top-heavy IMF (i.e., with a slope of $x=1.15$ as opposed to a standard $x=1.3$), 
and a lower fraction of binaries that explode as Type Ia supernovae appears to be 
required.  We also examine the rate of supernova explosions in the simulated galaxies. 
In early-type (non-star forming) galaxies, our predictions are also consistent with the 
observed SNe rates. However, in star-forming galaxies, a higher fraction of SN Ia 
binaries than in our preferred model is required to match the data. If, however,
we deviate from the classical model and introduce a population of SNe Ia with very short 
delay times, our models simultaneously produce a good match to the observed 
metallicities, abundance ratios and SN rates.

\end{abstract}

\begin{keywords}
galaxies: formation -- galaxies: chemical evolution
\end{keywords}

\section{Introduction}

The chemical properties and abundance ratios of galaxies provide important constraints 
on their formation histories. Galactic chemical evolution has been modelled in detail in 
the monolithic collapse scenario \citep[e.g.,][]{MG86,Francois04, Romano05,PM04,PM06}. 
These models have successfully described the abundance distributions in our Galaxy and 
other spiral discs, as well as the trends of metallicity and abundance ratios observed 
in early-type galaxies. In the last three decades, however, the paradigm of hierarchical 
assembly in a Cold Dark Matter (CDM) cosmology has revised the picture of how structure 
in the Universe forms and evolves. In this scenario, galaxies form when gas radiatively 
cools and condenses inside dark matter haloes, which themselves follow dissipationless 
gravitational collapse \citep{WR78,WF91}. The CDM picture has been successful at 
predicting many observed properties of galaxies, though many potential problems and open 
questions remain. It is therefore interesting to see whether chemical evolution models, 
when implemented within this modern cosmological context, are able to correctly predict 
the observed chemical properties of galaxies.

The semi-analytic approach provides a cosmological framework in which to study galaxy 
formation and chemical evolution in different environments, by following the merger 
history of dark matter haloes and the relevant physical processes such as gas cooling, 
star formation and feedback \citep[e.g.][]{WF91,KWG93,Cole94,Cole00,GalICS,sp,spf}. A 
major challenge for models of galaxy formation within the CDM picture arises from the 
mismatch between the {\em shape} of the mass function of the dark matter haloes and that 
of the baryonic condensations that we call galaxies \citep{WF91,KWG93,sp,benson:lf}. The 
CDM theory predicts a steeper slope for low-mass halos, and a more gradual drop-off in 
the abundance of high-mass halos than is seen in luminous galaxies, implying that the 
formation of stars must be inefficient in both low-mass and high-mass haloes 
\citep{Moster08}. However, the inclusion of physically motivated, if still {\em ad hoc}, 
feedback processes in the semi-analytic models can cure these discrepancies. The faint 
end of the luminosity function can be matched with a combination of supernova feedback 
and suppression of gas cooling in low mass haloes as a result of a photo-ionising 
background. At the bright end, heating by giant radio jets powered by accreting black 
holes has become a favored mechanism  for preventing over-cooling and quenching star 
formation in massive halos \citep{Croton06,Bower06,s08}. This latest generation of 
semi-analytic models (`SAMs') is successful at reproducing many properties of galaxies 
at the present and at high redshift, for example, the luminosity and stellar mass 
function of galaxies, color-magnitude or star formation rate vs. stellar mass 
distributions, relative numbers of early and late-type galaxies, gas fractions and size 
distributions of spiral galaxies, and the global star formation history \citep[e.g.][to 
name just a few]{Croton06,Bower06,Cattaneo06,deLucia06,s08,kimm:09,fontanot:09}.

The modelling of chemical enrichment of the galaxies and intergalactic (and 
intracluster) gas, however, has not been thoroughly developed in semi-analytic models, 
and to date most SAMs have only used the instantaneous recycling approximation (in 
essence only considering enrichment by type II supernovae) and trace only the total 
metal content. There are, however, a few models that have included a more refined 
treatment of the chemical enrichment. \citet{Thomas99} and \citet{TK99} were the first 
to include enrichment by type Ia supernovae (SNe Ia) in models with cosmologically 
motivated star formation histories.  However, rather than implementing the chemical 
evolution self-consistently within a semi-analytic model, they made use of star 
formation histories from the SAM and assumed a closed-box model (no gas inflows or 
outflows) for the chemical evolution. They calculated the evolution of [Fe/H] and 
[Mg/Fe] and found a decreasing trend of [Mg/Fe] with increasing galaxy luminosity, in 
stark disagreement with observations \citep[e.g.][]{WFG92,T00a,T00b,TMBO05}.

The first semi-analytic model to self-consistently track a variety of elements due to 
enrichment by SNe Ia and type II supernovae (SNe II) was that of 
\citet{Nagashima05a,Nagashima05b}. Among other things, they adopted a bimodal IMF 
described by a standard IMF for normal quiescent star formation in discs and an 
extremely flat `top-heavy' IMF during merger-driven starbursts. This model was motivated 
by the difficulty that semi-analytic models with a standard IMF experienced in 
reproducing the observed population of very luminous sub-mm galaxies at high redshift 
\citep{baugh:05}. However, the notion that early-type galaxies form their stars with an 
IMF flatter than standard is not new and has been proposed many times in the past as a 
plausible explanation for the abundance patterns in early-type galaxies and in the ICM 
of galaxy clusters \citep*[e.g.,][]{WFG92,MG95,GM97,TGB99}. The predictions of the 
Nagashima et al. model were in good agreement with the abundances of the intracluster 
medium (ICM) of galaxy clusters, matching the trend of individual elements (O, Fe, Mg, 
Si) and abundance ratios with ICM temperature. However, the same model failed to 
reproduce the trend of [$\alpha$/Fe] in early-type galaxies, where they found that the 
abundance ratio decreases with increasing galactic velocity dispersion, again in clear 
contradiction with observations. Very recently, \citet{Pipino08} have coupled galactic 
chemical evolution to the GalICS semi-analytic model \citep{GalICS}, and obtained 
results similar to those of \citet{Nagashima05b}.

In a simple closed-box picture, it is well-known that galaxies with short star formation 
timescales are expected to have enhanced [$\alpha$/Fe] ratios (because their enrichment 
is dominated by $\alpha$-rich Type II SNe), while galaxies with extended star formation 
histories tend to have lower [$\alpha$/Fe] 
\citep[e.g.][]{WFG92,TGB99,Thomas99,TK99,T00b}, because of the additional Fe contributed 
by delayed Type Ia enrichment. Therefore a possible interpretation of the difficulties 
that CDM-based galaxy formation models have experienced in reproducing the positive 
trend between mass, luminosity, or velocity dispersion and abundance ratio is related to 
the issue of so-called ``downsizing''. This refers to the variety of observational 
evidence that high-mass galaxies formed their stars early and over short timescales, 
while low-mass galaxies have more extended star formation histories \citep[see][for a 
summary]{fontanot:09}. Before the inclusion of AGN feedback or some other mechanism that 
quenches star formation in massive halos, CDM-based galaxy formation models predicted 
the opposite trend (massive galaxies continued to accrete gas and form stars until 
the present day, leading to extended star formation histories). It has been demonstrated 
that including radio-mode AGN feedback in semi-analytic models leads to a ``downsizing'' 
trend for star formation that is at least qualitatively in better agreement with 
observations \citep{deLucia06,s08,TS09,fontanot:09}. Therefore we expect that the 
new models might do better at reproducing the trend of [$\alpha$/Fe] with mass as the 
more massive galaxies will have shorter star formation timescales.

Clearly, observations of chemical abundances and abundance ratios in various phases 
(stellar, ISM, ICM) offer the opportunity to obtain strong constraints on galaxy 
formation histories and the physics that shapes them. However, in order to take 
advantage of these observations, it is necessary to implement detailed modeling of 
chemical evolution into a full modern SAM that includes the relevant physical processes 
(e.g. triggered star formation and morphological transformation of galaxies via mergers, 
the growth of supermassive black holes, and AGN feedback). In this work we incorporate 
detailed chemical evolution into the semi-analytic galaxy formation model of 
\citet{s08}, taking into account enrichment by SNe Ia, SNe II and long-lived stars, and 
abandoning the instantaneous recycling approximation by considering the finite lifetimes 
of stars of all masses. The delay in the metal enrichment by SNe Ia is calculated 
self-consistently according to the lifetimes of the progenitor stars. This is, to our 
knowledge, the first time that detailed chemical evolution has been included in a 
semi-analytic model with AGN feedback (both radio-mode heating and AGN-driven winds). 
The base model includes gas inflows due to radiative cooling of gas and outflows due to 
supernova and AGN-driven winds. We compute the abundances of many $\alpha$ and Fe-peak 
elements for early-type galaxies of different masses, exploring different IMF slopes and 
values for the fraction of binaries that yield a SN Ia event, and compare these with 
observations of abundances and abundance ratios for a sample of local early-type 
galaxies. We also calculate SNe rates using both the classical \citet{GR83} 
approach for type Ia SN and the more recent Delay-Time-Distribution (DTD) formalism 
\citep{greggio05}. Another improvement in the present work is our use of re-calibrated 
estimates for chemical abundances obtained from line-strengths in early type galaxies 
(see Appendix B for details).

The outline of the paper is as follows. In Section 2 we give an overview of the main 
ingredients of the semi-analytic model. In Section 3 we describe in detail the adopted 
treatment for the chemical evolution. In Section 4 we present our predictions and 
compare them with observations. In Section 5 we summarise our findings and present our 
conclusions. Two appendices describe the detailed implementation of the chemical 
evolution model and the data used in this paper.

\section{The semi-analytic model}

In this section we summarise the basic ingredients of the SAM used to model the 
formation and evolution of galaxies. These include the growth of structure of the dark 
matter component in a hierarchical clustering framework, radiative cooling of gas, star 
formation, supernova feedback, AGN feedback, galaxy merging within dark matter haloes, 
metal enrichment of the ISM and ICM, and the evolution of stellar populations. The 
reader is referred to \citet{sp}, \citet{spf} and especially \citet[hereafter S08]{s08} 
for a comprehensive and detailed description of the different prescriptions used in this 
semi-analytic model. In what follows we briefly sketch the modelling of the most 
important physical processes.

\subsection{Dark matter merger trees and galaxy merging}

The merging histories (or merger trees) of dark matter haloes are constructed 
based on the Extended Press-Schechter formalism using the method described in 
\citet{sk:99}, with improvements described in S08. Each branch in the tree 
represents a merger event, and, in order to make the process finite, the trees 
are followed down to a minimum progenitor mass of $10^{10}\,M_{\sun}$.

Whenever dark matter haloes merge, the central galaxy of the largest progenitor 
becomes the new central galaxy, and all others become `satellites'. Satellite 
galaxies may eventually merge with the central galaxy due to dynamical friction. 
To model the timescale of the merger process we use a variant of the 
Chandrasekhar formula from \citet{Boylan-Kolchin08}. Tidal stripping and 
destruction of satellites are also included as described in S08.

\subsection{Gas cooling, star formation and supernova feedback}

Before the Universe is reionised, each halo contains a mass of hot gas equal to 
the universal baryon fraction times the virial mass of the halo. After 
reionisation, the photo-ionising background can suppress the collapse of gas into 
low-mass halos.  We use the results of \citet{Gnedin00} and \citet{Kravtsov04} to 
model the fraction of baryons that can collapse into haloes of a given mass after 
reionisation.

When a dark matter halo collapses, or merges with a larger halo, the gas within 
it is shock-heated to the virial temperature of the halo, and gradually radiates 
and cools at a rate given by the cooling function. To calculate this function we 
use the metallicity-dependent radiative cooling curves of \citet{SD93}. A 
detailed description of how the cooling process is modelled can be found in S08. 
The rate at which gas can cool is given by:
\begin{equation}
\dot{m}_{\mathrm{cool}}=\frac{1}{2}m_{\mathrm{hot}}\frac{r_{\mathrm{cool}}}
{r_{\mathrm{vir}}}\frac{1}{t_{\mathrm{cool}}},
\end{equation}
where $m_{\mathrm{hot}}$ is the mass of the hot halo gas, $r_{\mathrm{vir}}$ is 
the virial radius of the dark matter halo, $r_{\mathrm{cool}}$ is the radius 
within which all of the gas can cool in a time $t_{\mathrm{cool}}$, which itself 
depends on density, metallicity and temperature. In our models, we assume that 
the cold gas is accreted only by the central galaxy of the halo, but in reality 
satellite galaxies should also receive some measure of new cold gas. This aspect 
of the modelling should be improved \citep[cf.][]{Pipino08} and for the present 
study we restrict our analysis to \emph{only} the central galaxy of each halo, 
except when otherwise stated.

When the gas cools we assume that it settles into a rotationally supported disc. 
The radial sizes of the discs are calculated according to the results described 
in \citet{somerville:08b}, and agree well with observed disc sizes to $z\sim 2$.

We model the star formation rate in quiescent discs with a recipe based on the 
empirical Schmidt-Kennicutt law \citep{Kennicutt89}:
\begin{eqnarray}
\dot{m}_{\star}&=&\frac{2\pi A_K \Sigma_0^{N_K}r_{\mathrm{gas}}^2}{N_K^2}
\nonumber \\
& & \times \left[1-\left(1+\frac{N_K r_{\mathrm{crit}}}{r_{\mathrm{gas}}}\right)
  \exp(-N_K r_{\mathrm{crit}}/r_{\mathrm{gas}})\right] \label{sfrec}
\end{eqnarray}
where $\dot{m}_{\star}$ is the star formation rate, $\Sigma_0\equiv 
m_{\mathrm{cold}}/(2\pi r_{\mathrm{gas}}^2)$ is the average surface density of 
the cold gas, $r_{\mathrm{gas}}$ is the scale-length of the gaseous disc (assumed 
to be an exponential disc with its scale-length proportional to that of the 
stellar disc), $r_{\mathrm{crit}}$ is the radius at which the gas reaches the 
critical surface density threshold for star formation ($\Sigma_{\mathrm{crit}}$), 
and $A_K$ and $N_K$ are the normalisation and slope of the SFR law. We adopt the values $A_K=8.35\times 10^{-5}$, $N_K=1.4$ and $\Sigma_{\mathrm{crit}}=6\, 
M_{\sun}{\mathrm{pc}^{-2}}$, as in S08.

Galaxy mergers in the SAM trigger enhanced episodes of star formation. The burst 
is modelled by two parameters, the time-scale and the efficiency of the burst. 
The time-scale is a function of the virial velocity of the progenitor galaxies, 
the equation of state of the gas, the cold gas fraction in the discs, and the 
redshift \citep{robertson:06}. The efficiency, which is defined as the fraction 
of the cold gas reservoir (of both galaxies) that is turned into stars during the 
burst, is assumed to be a power-law function of the mass ratio of the merging 
galaxies, and the exponent of the power-law depends on the galaxy morphology 
\citep{cox:08}. The collisional starburst occurs in addition to any ongoing 
`normal' quiescent star formation, which continues uninterrupted through the 
merger but is usually insignificant in comparison to the burst. Any new stars 
formed during the burst are always placed in the bulge component of the resulting 
galaxy.

As supernovae occur, they inject energy into the ISM and reheat the cold gas, 
which is then expelled from the disc and incorporated into the hot halo gas where 
it can cool again. The rate of reheating by SNe is given by
\begin{equation}
\dot{m}_{\mathrm{rh}}=\epsilon_0^{SN}\left(\frac{V_{\mathrm{disc}}}{200\,
\mathrm{km/s}}\right)^{-\alpha_{\mathrm{rh}}} \dot{m}_{\star}, \label{reheat}
\end{equation}
where $\epsilon_0^{SN}$ and $\alpha_{\mathrm{rh}}$ are free parameters. The 
circular velocity of the disc $V_{\mathrm{disc}}$ is taken to be equal to the 
maximum rotational velocity of the dark matter halo. Some fraction of the 
reheated gas can also be ejected from the halo entirely into the diffuse 
Intergalactic Medium. This fraction is described by:
\begin{equation}
f_{\mathrm{eject}}(V_{\mathrm{vir}})=[1.0+(V_{\mathrm{vir}}/V_{\mathrm{eject}})
^{\alpha_{\mathrm{eject}}}]^{-1}, \label{expul}
\end{equation}
where $\alpha_{\mathrm{eject}} = 6$ and $V_{\mathrm{eject}}$ is a free parameter 
in the range $\simeq 100-150$ km/s. This ejected gas is allowed to re-collapse 
into the halo at later times and once again becomes available for cooling.

\subsection{Formation of Spheroids}

In most semi-analytic models each merger is classified as `major' or `minor' 
depending on whether the ratio of the smaller to the larger galaxies' baryonic 
masses is greater than or less than the parameter $f_{\rmn{ellip}} \sim 0.25$, 
respectively. The usual assumption is then that, in a major merger, the bulge and 
disc stars of both progenitor galaxies, as well as the stars formed in the merger 
driven starburst (see below), are transferred to the bulge component of the 
resulting galaxy. In a minor merger, all the pre-existing stars of the smaller 
galaxy end up in the disc of the post-merger galaxy, and all the newly formed 
stars are placed in bulge. We follow a similar practise here, but instead of 
using a sharp threshold to define major or minor mergers, we use a more gradual 
transition function. In detail, when two galaxies with bulge masses $B_1$ and 
$B_2$, and disc masses $D_1$ and $D_2$ merge, the resulting galaxy has a bulge 
mass $B_{\rmn{new}}=B_1 + B_2 + f_{\rmn{sph}}(D_1 + D_2)$ and a disc mass
$D_{\rmn{new}}=(1-f_{\rmn{sph}})(D_1+D_2)$. The value $f_{\rmn{sph}}$ is a 
continuous function of the \textit{total} mass ratio (baryons and dark matter) in 
the \textit{central} parts of the galaxy (see S08).

\subsection{Black Hole Growth and AGN Feedback}

The models of S08 also track the growth of super-massive black holes and the 
energy they release. Each top-level DM halo is seeded with a $\sim 100$ $\msun$ 
black hole, and these black holes are able to grow via two different accretion 
modes. The first accretion mode is fuelled by cold gas that is driven into the 
nucleus of the galaxy by mergers. This mode is radiatively efficient, and the 
accretion rates are close to the Eddington limit. Because this accretion mode is 
associated with optically bright classical quasars and AGN, it is referred to as 
`bright mode' or `quasar mode' accretion. The second mode is fuelled by hot gas 
in a quasi-hydrostatic halo, and the accretion rate is modelled via the 
Bondi-Hoyle approximation. Accretion rates in this mode are significantly 
sub-Eddington ($\sim 10^{-4}$ to $10^{-3}$ times the Eddington rate) and the 
accretion is assumed to be radiatively inefficient. This mode is, however, 
associated with the production of giant radio jets, and is therefore referred to 
as the `radio mode'.

Energy released during `bright mode' activity can couple with the cold gas in the 
galaxy via radiation pressure, driving galactic scale winds that can eject cold 
gas from the galaxy. The mass outflow rate due to the AGN driven wind is modelled 
by the following formula:
\begin{equation}
\dot{m}_{\mathrm{agn}}=\epsilon_{\mathrm{wind}}\eta_{\mathrm{rad}}\frac{c}
{V_{\mathrm{esc}}}\dot{m}_{\mathrm{acc}},
\end{equation}
where $\epsilon_{\mathrm{wind}}$ is the effective coupling efficiency, 
$V_{\mathrm{esc}}$ is the escape velocity of the galaxy and 
$\dot{m}_{\mathrm{acc}}$ is the accretion rate of mass onto the black hole.

The radio jets produced by `radio mode' activity are assumed to inject thermal 
energy into the hot halo gas, partly or completely offsetting the cooling flow. 
This process is responsible for quenching the star formation in massive galaxies 
(which contain massive black holes) and solves the `over-cooling problem' that 
plagued CDM-based galaxy formation models for many years.

\subsection{Stellar Population Synthesis and Dust}

In order to compare the luminosities and colours of the galaxies in the 
simulations with real observations, we convolve the star formation and chemical 
enrichment history of each galaxy with the multi-metallicity simple stellar 
population (SSP) models of \citet{BC03}. We use the models based on the 
Padova1994 \citep{Padova} isochrones with a \citet{Chabrier01} IMF.

We also model the effects of dust extinction. Based on the model of 
\citet{charlot_fall:00}, we consider extinction due to two components, one due to 
the diffuse dust in the disc and another associated with the dense `birth clouds' 
surrounding young star forming regions. The $V$-band, face-on extinction optical 
depth of the diffuse dust is given by
\begin{equation}
\tau_{V,0}\propto\tau_{\mathrm{dust,0}}Z_{\mathrm{cold}}m_{\mathrm{cold}}/
(r_{\mathrm{gas}})^2,
\end{equation}
where $\tau_{\mathrm{dust,0}}$ is a free parameter, $Z_{\mathrm{cold}}$ is the 
metallicity of the cold gas, $m_{\mathrm{cold}}$ is the mass of the cold gas in 
the disc, and $r_{\mathrm{gas}}$ is the radius of the cold gas disc. To compute 
the actual extinction we assign each galaxy a random inclination and use a 
standard `slab' model. Additionally, stars younger than $10^7$ yr are enshrouded 
in a cloud of dust with optical depth 
$\tau_{\mathrm{BC,V}}=\mu_{\mathrm{BC}}\tau_{V,0}$, where $\mu_{\mathrm{BC}}=3$. 
Finally, to extend the extinction correction to other wavebands, we assume a 
Galactic attenuation curve \citep{cardelli:89} for the diffuse dust component and 
a power-law extinction curve $A_{\lambda}\propto(\lambda/5500\AA)^n$, with 
$n=0.7$, for the birth clouds.

\subsection{Cosmological and Galaxy Formation Parameters}

We adopt a flat $\Lambda$CDM cosmology with $\Omega_{0}=0.2383,\ 
\Omega_{\Lambda}=0.7617,\ h\equiv H_{0}/(100\,\mathrm{km\ s}^{-1} 
\mathrm{Mpc}^{-1})=0.732,\ \sigma_{8}=0.761$, and a cosmic baryon fraction of 
$f_{b}=0.1746$, following the results of \citet{WMAP3}.  We adopt these 
parameters for consistency with the published models of S08, but find that we 
obtain nearly identical results with the updated values of the cosmological 
parameters from \citet{komatsu09}.

We leave the values of the free parameters associated with the galaxy formation 
models fixed to the fiducial values given in S08. These values were chosen by 
requiring that the models reproduced key observations of nearby galaxies, such as 
the $z\sim 0$ stellar mass function, and gas fractions and star formation rates 
as a function of stellar mass. These models have also been shown to produce 
reasonable agreement with observed local galaxy colour distributions in 
\citet{kimm:09}, and with observed stellar mass functions and star formation 
rates at high redshift \citep[$0<z<4$;][]{fontanot:09}. In \S~\ref{sec:impact} we 
check that our new models, with the updated treatment of chemical evolution 
modelling, still reproduce the key observational quantities with the same values 
of the free parameters.

\section{Galactic Chemical Evolution}

In S08, the production of metals was tracked using a simple approach commonly 
adopted in semi-analytic models \cite[see e.g.][]{sp,Cole00,deLucia04,kang05}. In 
a given time-step, when we create a parcel of new stars ${\rm d}m_*$, we also 
create a mass of metals ${\rm d}M_Z = y \, {\rm d}m_*$, which we assume to be 
instantaneously mixed with the cold gas in the disc. The yield $y$ is assumed to 
be constant, and is treated as a free parameter.\footnote{The yield parameter $y$ 
should not take arbitrary values since it is constrained by the IMF and the 
theory of stellar nucleosynthesis (albeit subject to the uncertainties in both). 
However, it is treated in most SAMs as a free parameter.} We 
track the mean metallicity of the cold gas $Z_{\rm cold}$, and when we create a 
new parcel of stars they are assumed to have the same metallicity as the mean 
metallicity of the cold gas in that time-step. Supernova feedback ejects metals 
from the disc, along with cold gas. These metals are either mixed with the hot 
gas in the halo, or ejected from the halo into the `diffuse' Intergalactic Medium 
(IGM), in the same proportion as the reheated cold gas. The ejected metals in the 
`diffuse gas' reservoir are also re-accreted into the halo in the same manner as 
the gas.

In the present study, we discard the instantaneous recycling approximation and 
allow the ISM to be enriched by the products of type Ia and type II supernovae on 
their own timescales. Consequently, we now track individual elements, and not 
just the total metal content. The integrated ejecta of each element is not a free
parameter, but instead is calculated according to theoretical yields and the 
star-formation histories provided by the SAM. In the next subsection we describe 
the implementation of the new chemical evolution model in detail.

\subsection{Basic equations of the GCE}

For the purposes of tracing the enrichment of the ISM, we still model our 
galaxies as a single zone with instantaneous mixing of gas. We assume that newly 
produced metals are deposited into the cold gas, and may subsequently be ejected 
from the galaxy and mixed with the hot halo gas (or ejected from the halo 
altogether) according to the feedback model described above. The metallicity of 
each new batch of stars equals that of the cold gas at the moment of formation. 
In this context, the evolution of the abundance of metals in the cold gas is 
given by 
\begin{equation}
  \dot{G}_{Z}(t)=-\psi(t)Z(t)+e_{Z}(t)+
  \left[\dot{G}_{Z}(t)\right]_{\mathrm{inflow}}
  -\left[\dot{G}_{Z}(t)\right]_{\mathrm{outflow}}, 
  \label{basic}
\end{equation}
where $G(t)$ is the total mass of gas and $Z(t)$ is the mass-weighted metal 
abundance, $G_Z(t)=G(t)Z(t)$ is the mass of gas in the form of metals, $\psi(t)$ 
is the star formation rate and $\psi(t)Z(t)$ represents the rate at which metals 
are depleted from the ISM by star formation, $e_Z(t)$ is the rate of ejecta of 
enriched material by dying stars (integrated over stellar mass), and the last two 
terms represent the infall of cooled halo gas into the galaxy and the outflow of 
reheated gas from the galaxy. Here we refer generally to `metals' for simplicity, 
but in fact we apply this equation to each individual element by considering the 
abundance $Z_i$ of a given element $i$ instead of the total metallicity $Z$. For 
comprehensive reviews of Eq.~\ref{basic}, we direct the reader to 
\citet{Tinsley80} and \citet{Pagel97}. The modelling of the star formation rate, 
the inflow rate (cooling flows) and the outflow rate (supernovae and AGN driven 
galactic winds) have already been sketched in the previous section. The different 
prescriptions shown before relate to the terms in Eq.~\ref{basic} in the 
following way:
\begin{eqnarray}
\psi(t)Z(t)&=&\dot{m}_{\star}Z_c; \nonumber \\
\left[\dot{G}_{Z}(t)\right]_{\mathrm{inflow}}&=&\dot{m}_{\mathrm{cool}}Z_h;
 \nonumber \\
\left[\dot{G}_{Z}(t)\right]_{\mathrm{outflow}}&=&(\dot{m}_{\mathrm{rh}} + 
\dot{m}_{\mathrm{agn}}) Z_c; 
\end{eqnarray}
where $Z_c$ and $Z_h$ are the abundances of the cold ISM gas and the hot halo 
gas, respectively.

In most SAMs previous to this work, chemical evolution was traced in a very 
simple manner by assuming a constant `effective yield', or mean mass of metals 
produced per mass of stars, and the value of this effective yield was treated as 
a free parameter. In the models presented here, we have implemented detailed 
calculations for the production of heavy elements and the chemical enrichment of 
the ISM (the second term in Eq.~\ref{basic}) in a similar fashion to the models 
of \citet{MG95}, \citet{Timmes95} and \citet{PM04}. In this framework, not only 
do we trace the evolution of the total metallicity, but we also track the 
distinct elements as well. At this moment we can follow the evolution of the 
abundances of 19 different elements, but here we will only discuss 
$\alpha$-elements and Fe. By $\alpha$-elements we mean the composite abundance of 
N, Na, Ne, Mg, Si and S.  At any given time, the rate at which an element $i$ 
restored into the interstellar medium is calculated according to the following 
formula
\newpage
\begin{eqnarray}
e_{Z_i}&=&\int_{M_L}^{M_{B_m}}\!\psi(t-\tau_M)Q_{mi}(t-\tau_M)\phi(M)\,
{\mathrm{d}}M \nonumber \\
&+&A\int_{M_{B_m}}^{M_{B_M}}\!\phi(M_B) \nonumber \\
& & \times \left[\int_{\mu_{\mathrm{min}}}^{0.5}\!f(\mu)\psi(t-\tau_{M_2})Q_{mi}
(t-\tau_{M_2})\mathrm{d}\mu\right]{\mathrm{d}}M_B \nonumber \\
&+&(1-A)\int_{M_{B_m}}^{M_{B_M}}\!\psi(t-\tau_M)Q_{mi}(t-\tau_M)\phi(M)\,
{\mathrm{d}}M \nonumber \\
&+&\int_{M_{B_M}}^{M_U}\!\psi(t-\tau_M)Q_{mi}(t-\tau_M)\phi(M)\,{\mathrm{d}}M
\label{ejec}
\end{eqnarray}
where $M_L$ and $M_U$ are the lower and upper masses of stars formed, $\psi(t)$ 
is the star formation rate as before, $\phi(M)$ is the initial mass function 
(IMF), $f(\mu)$ is the distribution function for the mass of the secondary star 
in a binary pair ($\mu=M_2/M_B$), $\tau_M$ is the lifetime of a star of mass $M$, 
and $Q_{mi}(t)$ represent the fractional mass of an element $i$ restored by a 
star of mass $M$ in the form of both newly synthesised and unprocessed material. 
Although not explicitly dependent on time, the quantities $Q_{mi}$ depend on 
metallicity, which of course evolves in time.
 
Each of the integrals in Eq.~\ref{ejec} represents the contribution to the 
enrichment by stars in different mass ranges. The first integral indicates the 
contribution of single stars with masses between $M_L=0.8M_{\sun}$ (the minimum 
mass which can restore gas to the ISM within a Hubble time) and 
$M_{B_m}=3M_{\sun}$ (the minimum mass of a binary system which can give rise to a 
SNe Ia event). These stars eject their chemical by-products through stellar winds 
and end their lives as white dwarfs. The second integral refers to the 
contribution from type Ia SNe, assuming that these events originate from C-O 
white dwarfs in binary systems exploding by C-deflagration after reaching the 
Chandrasekhar mass. This implies a maximum primary mass of $8M_{\sun}$, and 
therefore $M_{B_M}=16M_{\sun}$ and $\mu_{\mathrm{min}}=\mathrm{max}
\left\lbrace\frac{M_L}{M_B},\frac{M_B-8M_{\sun}}{M_B}\right\rbrace$. The 
parameter $A$ represents the fraction of binary systems with total mass in the 
appropriate range that actually give rise to a SNe Ia event. In essence, $A$ is a 
free parameter. Chemical evolution models of the Milky Way constrain the value of 
this parameter to around $A\sim0.04-0.05$ by ensuring compatibility with the 
observed present-day rate of SNe I and SNe II in our galaxy \citep{Francois04}. 
However, this value results in an unacceptably high abundance of Fe and Fe-peak 
elements in our models. Therefore, we allow this parameter to take different 
values (0.015, 0.02, 0.03, 0.05) and constrain it \emph{a posteriori} by 
comparison with abundance ratios and SNe rates \citep[see also the discussion 
in][]{dePlaa07}. The distribution function of the secondary mass fraction is 
assumed to follow the law
\begin{equation}
f(\mu)=2^{1+\gamma}(1+\gamma)\mu^{\gamma},
\end{equation}
with $\gamma=2$. A complete description of all the quantities involved in the 
computation of the SNe I rate can be found in \citet{GR83} and \citet{MG86}. Note 
that in this scheme, it is the mass of the secondary star ($M_2$) that sets the 
clock for the explosion. This {\it implies} a specific delay time distribution 
(DTD) for the explosions that may not represent reality (see 
\S~\ref{sec:res}). The third integral represents the mass restored by stars in 
the mass range 3--$16\,M_{\sun}$ which are either single, or, if binaries, do not 
produce a SN I event. These stars end their lives as white dwarfs 
($M<8\,M_{\sun}$) or as SNe II ($M>8\,M_{\sun}$). Finally, the last term 
represents the contribution of short-lived massive stars ($M>16\,M_{\sun}$) that 
explode as SNe II. The fact that we take into account the lifetimes of the stars 
implicitly involves a time delay for the different enrichment modes (AGB stellar 
winds, SNe I and SNe II) since the integrands in Eq.~\ref{ejec} are by definition 
zero whenever $t<\tau_M$.

Before applying the GCE to the semi-analytic model, we tested the chemical 
evolution code separately. We ran simulations using only the chemical evolution 
algorithm and compared the results with simple models with analytic solutions, 
i.e., a closed box with either constant and continuous star formation or a single 
initial burst of star formation. We also compared our results with the output 
from other well-tested models, namely that of \citet{FG03}. Only after achieving 
satisfactory agreement in these tests did we proceed to implement the chemical 
evolution into the SAM.

\subsection{Ingredients of the GCE}

In the previous section, we introduced several fundamental quantities that 
determine the chemical enrichment of a galaxy's cold gas reservoir and described 
them in a qualitative manner. In what follows, we will point the reader to the 
different studies that quantify the ingredients of this model, and the values 
adopted for the simulations presented in the next section. We discuss the 
implementation in Appendix A.

\subsubsection{Initial Mass Function}

The stellar initial mass function that we use is based on the parameterization of 
\citet{Chabrier03}:
\begin{equation}
\phi(m)=\left\lbrace \begin{array}{ll}
\mathcal{A}\,\mathrm{e}^{-(\log m-\log m_c)^2/2\,\sigma^2} & \mbox
{if $m < 1\,M_{\sun}$} \\
\mathcal{B}\, m^{-x} & \mbox{if $m > 1\,M_{\sun}$} 
\end{array}\right. \label{IMF}
\end{equation}
where in the standard Chabrier IMF, $x = 1.3$, $\sigma=0.69$, $m_c=0.079\,M_{\sun}$, and the proportionality constants take the values 
${\mathcal{A}} = 0.9098$ and ${\mathcal{B}} = 0.2539$ after normalisation in the 
mass interval $0.1-40\,M_{\sun}$. This IMF differs somewhat from the standard 
power laws of \citet{SalpeterIMF} and \citet{KroupaIMF} often used in the 
literature. The reason for this choice is consistency with the stellar population 
synthesis models that are used to predict magnitudes and colors in the 
simulations, since those models use the Chabrier IMF \citep{BC03}. Note that this 
expression is for the IMF by mass.

We show below that the results with this IMF were not entirely satisfactory.  We therefore explored different slopes ($x=1.2,1.1,1.0$) and upper mass limits 
($M_U=60,100,120\,M_{\sun}$). The values of the normalisation constants 
${\mathcal{A}}$ and ${\mathcal{B}}$ for different values of $x$ and $M_U$ 
are computed by requiring that
\begin{equation}
\int_{M_L}^{M_U}\phi(m)\,{\mathrm{d}}m=1.
\end{equation}
We note that these alternate values for the slope are within the observational 
uncertainties \citep{Chabrier03}, and certainly do not represent a radical 
departure from the observed local IMF.

\subsubsection{Stellar lifetimes}

The adopted relation between the evolutionary lifetimes and the stellar mass is 
that of Padova tracks for solar metallicity \citep{Padova}. In principle the 
lifetimes depend not only on mass but also on metallicity. Nevertheless the 
difference in stellar ages for different metal abundances is smaller than the age 
binning in the grid that stores the star formation history of the galaxies in our 
simulations (see Appendix A).

\subsubsection{Stellar yields}

Stellar yields are the amount of material that a star can produce and eject into 
the ISM in the form of a given element, and are clearly one of the most important 
ingredients in any chemical evolution model. These yields are the quantities 
$Q_{mi}$ in the equations above. In this work we adopt different nucleosynthesis 
prescriptions for stars in the different mass ranges.
\begin{description}
\item{Low and intermediate mass stars} ($0.8<M/M_{\sun}<8$) produce He, C, N, and 
  heavy s-process elements\footnote{At this point we do not trace s-process or 
  r-process elements, but they will be included in future versions of the code.}, 
  which they eject during the formation of a planetary nebulae. The yields that 
  we adopt are from \citet[hereafter KL07]{KL07}.
\item{Massive stars} ($M>8M_{\sun}$) produce mainly $\alpha-$elements (O, Na, Ne, 
  Mg, Si, S, Ca), some Fe-peak elements, light s-process elements and r-process 
  elements. They explode as core-collapse Type II SNe. We adopt the yields from 
  \citet[hereafter WW95]{WW95}. Note that the upper mass limit in this study is 
  $40\,M_{\sun}$.
\item{Type Ia SNe} are assumed to be C-O white dwarfs in binary systems, 
  exploding by C-deflagration after reaching the Chandrasekhar mass via accretion 
  of material from the companion star. They mainly produce Fe and Fe-peak 
  elements. The yields we adopt are from \citet[hereafter N97]{Nomoto97}, model 
  W7. When calculating the contribution of SNe Ia, we assume that the primary 
  star also enriches the medium as a normal AGB prior to the SN event.
\end{description}

Except for the SN Ia yields, which are given only for solar metallicity, we use 
metallicity dependent yields, namely those tabulated for 
$Z=0.0002,\,0.004,\,0.02$, interpolating when necessary but never extrapolating. 
Whenever the metallicity falls below or above the limiting values, we use the 
yields corresponding to the minimum or maximum $Z$ respectively.

Note that unlike in some recent studies of galactic chemical evolution
\citep{Francois04,PM04,PM06,Nagashima05a,Nagashima05b} we do not alter the yields 
in any way in our standard model. We want to see if we can fit the data with as 
few degrees of freedom as possible.

\subsection {Delay Time Distribution formulation for SNe Ia}

As mentioned before, and shown in \S~\ref{sec:res}, the SN Ia
  model described in the previous subsection does not seem to be the
  best representation of this phenomenon. In order to test other
  models, we implemented the delay-time-distribution (DTD) formalism
  developed by \citet{greggio05}. In this scenario, the SN Ia rate is
  described by
\begin{equation}
R_{Ia}(t)=k_{\alpha}\int_{\tau_i}^{{\rm min}(t,\tau_x)}\!A(t-\tau)\psi(t-\tau)DTD(\tau)
 \,{\mathrm{d}}\tau
\end{equation}
where $\psi(t)$ is the star formation rate, $\tau_i$ is the
minimum delay time for the SNIa events which we assume to be equal to
the lifetime of an $8\,M_{\sun}$ star, $\tau_x$ is the maximum
delay and equal to the lifetime of a $0.8\,M_{\sun}$ star, and
$k_{\alpha}$ is the number of stars per unit mass in a stellar
generation defined by
\begin{equation}
k_{\alpha}=\int_{M_L}^{M_U}\!\phi(m)m^{-1}\,\mathrm{d}m.
\end{equation}
Finally, $A(t-\tau)$ is the fraction of binary systems which give rise
to Type Ia SNe and may, in principle, evolve in time, but here we will
assume it to be constant. It should be noted that in this case,
$A(t-\tau)$ is the fraction relative to the full mass range defined by
the IMF $(M_L-M_U)$ and not only the mass range $3-16\,M_{\sun}$ as
before. To ease the comparison with the previous model we will define
$A(t-\tau)=A\,f_{3-16}$, where $f_{3-16}$ is the fraction of stars in
the $3-16\,M_{\sun}$ mass range (defined by the IMF) and $A$ has the
usual meaning.

\begin{figure}
\includegraphics[width=85mm]{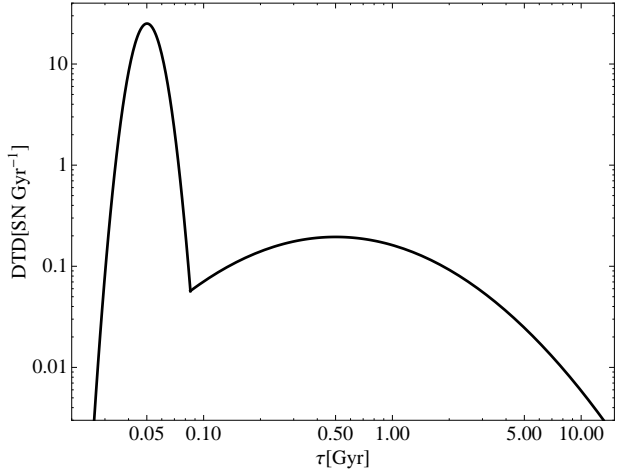}
\caption{The normalized DTD proposed by \citet{MDP06}, and adopted in
  this work.}
\label{DTDp}
\end{figure}

This formulation allows for different SNe Ia models depending on the
DTD($\tau$) used. In particular we have chosen the distribution
favoured by \citet{MDP06}, and parametrized it in similar way as
\citet{matteucci06}:
\begin{equation}
\log DTD(\tau)=\left\lbrace \begin{array}{ll}
\!1.4-50(\log \tau +1.3)^2 & \! \mbox{if $\tau < \tau_o$} \\
\!-0.71-0.9(\log \tau +0.3)^2 & \! \mbox{if $\tau > \tau_o$} \\
\end{array}\right. \label{DTDf}
\end{equation}
where delay time $\tau$ is in Gyrs, $\tau_o=0.0851$Gyr, and the
distribution function is normalized so that
\begin{equation}
\int_{\tau_i}^{\tau_x}\!DTD(\tau)\,{\mathrm{d}}\tau = 1
\end{equation}
In Figure~\ref{DTDp} we show the behaviour of the implemented
DTD. Note that for a single burst of star formation, about half of all
SN Ia explosions occur within the first 100 Myr.

\section{Results}

In this section we present the first results of our model and compare them with 
stellar population studies of a variety of early-type galaxies in the local 
universe. For this purpose, we ran simulations for a grid of dark matter haloes 
of different masses, ranging from $10^{11}$ to $10^{13}\,M_{\sun}$, using both 
the original (instantaneous recycling) and new (full GCE) version of the 
semi-analytic code. As mentioned before, we limit our analysis to the central 
galaxies of each DM halo. We selected early-type galaxies from our simulations 
according to the ratio of their bulge-to-total luminosity in the B-band; namely, 
we consider a galaxy to be an early-type when this ratio is larger than 0.4047 
\citep{SdV86}. This selection encompasses both elliptical and S0 galaxies. Unless 
otherwise noted, the model results presented in this section are always for 
central early-type galaxies. We begin with a comparison of the results of our new 
model with those of the S08 SAM. We then proceed to compare the predictions of 
the new model with observations. For the purpose of this comparison, we will only 
show model galaxies with masses above $10^9\,M_{\sun}$ since the formation 
history for galaxies below this mass cannot be accurately resolved given the mass 
resolution of the dark matter trees (see \S 2.1).

\subsection{Impact of the new GCE modelling in galaxy observables}
\label{sec:impact}

The new chemical evolution modelling affects the physics in the SAM in at least 
three ways: changing the metallicity of the hot gas changes the cooling rates; 
changing the metallicity of the cold gas changes the amount of dust, and 
therefore observed colours and magnitudes; and the metallicities themselves and 
their evolution change as well. The original version of the S08 SAM did very well 
at reproducing several key properties of galaxies and it is important to verify 
that this is still the case after implementing the detailed chemical evolution 
model. For this purpose we will compare simulations with three different 
`flavors' for the models: the {\it original} SAM from S08, a SAM+GCE with {\it 
standard} parameters (i.e. $x=1.3$ and $A=0.05$), and a SAM+GCE with parameters 
that best fit ({\it best-fitting}) the abundance ratios and metallicities of 
observed galaxies ($x=1.1$ and $A=0.015$, see \S ~\ref{sec:res}).

\begin{figure}
\includegraphics[width=85mm]{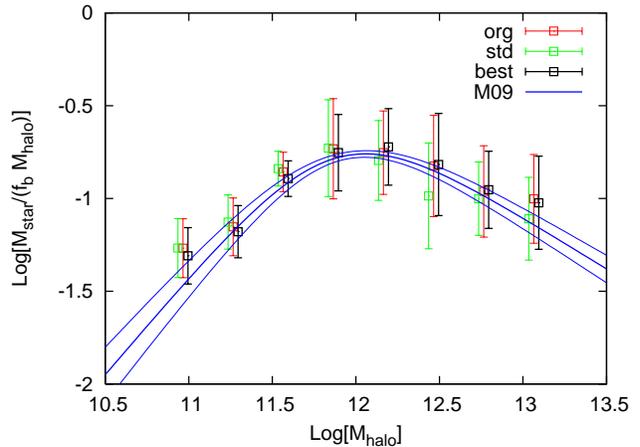}
\caption{The fraction of baryons in the form of stars as a function of halo mass 
for model galaxies in the SAM+GCE with the {\it best-fitting} parameters ($x=1.1$ 
\& $A=0.015$, black squares), with {\it standard} parameters ($x = 1.3$ \& 
$A=0.05$, green squares), and in the {\it original} SAM without GCE (red 
squares). The blue lines mark the empirical relation, and 1-$\sigma$ 
uncertainties, derived by \citet{Moster08}.}
\label{bar-frac}
\end{figure}

A fundamental feature that any good galaxy formation model must reproduce is the 
luminosity function or stellar mass function of galaxies. Given the distribution 
function of dark matter halos and sub-halos predicted by CDM, the relationship 
between stellar mass and halo mass implies a specific stellar mass function. The 
required stellar mass to halo mass relationship, in the form of the fraction of 
baryons in the halo that are converted to stars in a galaxy, has been derived by 
\citet{wang06} and \citet{Moster08}. In Figure~\ref{bar-frac} we show the stellar 
mass fraction as a function of halo mass for the SAM with and without GCE and 
compared with the empirical relation obtained by \citet{Moster08}. The agreement 
between all `flavors' of the models (S08, standard parameters, best-fitting) 
and the observations is excellent. This implies that the stellar mass function in 
our new models will be nearly identical to that presented in S08, which was shown 
to agree well with observations.

\begin{figure}
\includegraphics[width=85mm]{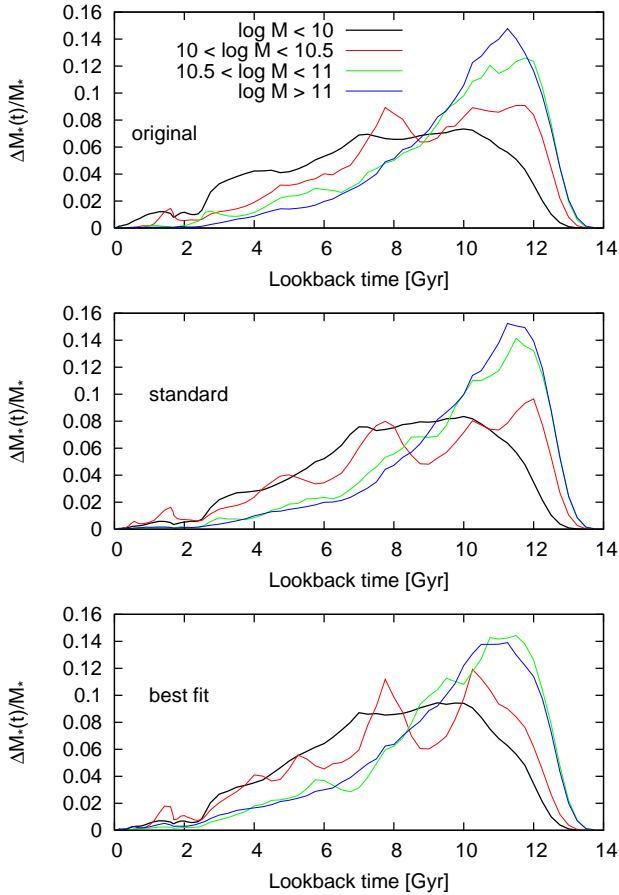}
\caption{Smoothed average star-formation histories for model early-type galaxies 
in (from bottom to top) the SAM+GCE with the {\it best-fitting} parameters 
($x=1.1$ \& $A=0.015$), with {\it standard} parameters ($x = 1.3$ \& $A=0.05$), 
and in the {\it original} SAM without GCE, binned by the stellar mass of the 
galaxy at $z=0$.}
\label{sfh}
\end{figure}

In Figure~\ref{sfh}, we show the average star formation histories (SFH) of our 
model early-type galaxies, for galaxies with different present-day stellar 
masses. As pointed out in previous studies \citep[see, e.g., \citealt{deLucia06}; 
S08;][hereafter TS09]{TS09}, SAMs with AGN feedback do qualitatively reproduce a 
downsizing-like trend (i.e., the higher mass galaxies have shorter SF timescales 
and less extended SFH). This is still true in our new models.

\begin{figure}
\includegraphics[width=85mm]{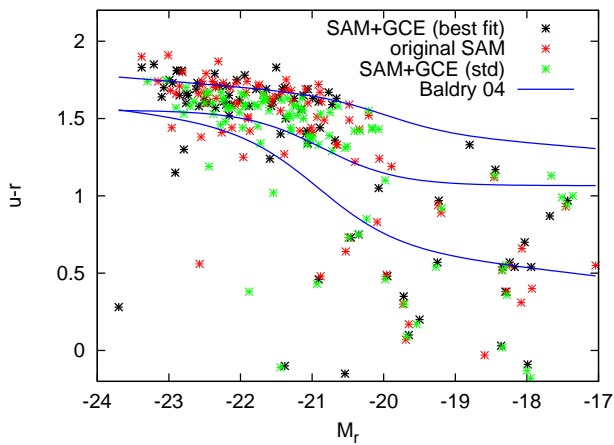}
\caption{Colour-magnitude diagram (CMD) for model early-type galaxies in the 
SAM+GCE with the {\it best-fitting} parameters ($x = 1.1$ \& $A=0.015$, black 
stars), with {\it standard} parameters ($x = 1.3$ \& $A=0.05$, green stars), and 
in the {\it original} SAM without GCE (red stars). The blue lines mark, from top 
to bottom, the locus of the red sequence, the green valley and the blue cloud 
from \citet{Baldry04}.}
\label{cmd}
\end{figure}

Another very important observational quantity to reproduce is the
colour-magnitude diagram (CMD) of galaxies. In Figure~\ref{cmd} we
show magnitudes and colours of early-type galaxies in the SAM with and
without GCE.  Here there is one caveat regarding the calculation of
the galaxy luminosities.  The stellar population models that we use to
predict the colours and magnitudes \citep{BC03} use a fixed standard
Chabrier IMF, while we allow the slope of this IMF to change when
calculating the chemical evolution. Although this is not
self-consistent, such a minor change in the IMF parameters should not
significantly affect the predicted colours or magnitudes since
  the early-type galaxies studied here are dominated by old
  populations for which high-mass stars are of little importance
(C. Conroy, private communication). We divide the CMD into red and
blue regions using the magnitude-dependent cut of
\citet{Baldry04}. The galaxies form two clear groups: the majority in
a bright red sequence and a few in a fainter blue cloud. The
``original'' and ``best fit'' models agree quite well with the
observed CMD, while the luminous galaxies in the ``standard model''
are slightly too blue.

\begin{figure}
\includegraphics[width=85mm]{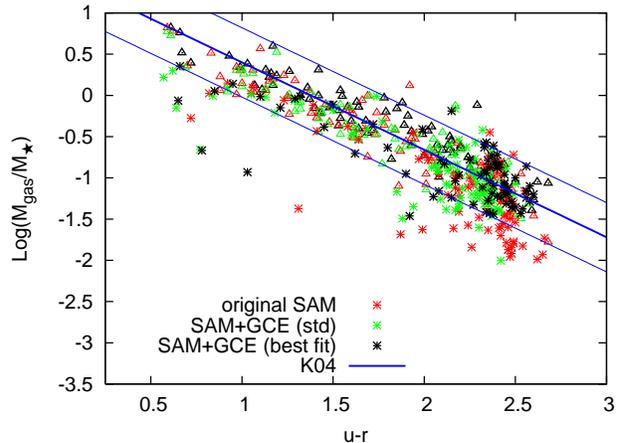}
\caption{The gas fraction for model galaxies in the SAM+GCE with the best fitting 
parameters ($x = 1.1$ \& $A=0.015$, black symbols), with standard parameters 
($x=1.3$ \& $A=0.05$, green symbols), and in the original SAM without GCE (red 
symbols) as a function of $u-r$ colour. Stars and triangles depict early-type and 
disc galaxies, respectively. The thick blue line marks the median of the sample 
from \citet{Kannappan:04}, and the thin lines mark the $1-\sigma$ deviation.}
\label{gas-frac2}
\end{figure}

In the previous version of the SAM, the fraction of cold gas relative to stars in 
galactic discs at the present time was used to calibrate the models by comparison 
with the observational estimates of \citet{Bell03} for morphologically late-type 
galaxies (see Figure 5 in S08). This property is well reproduced by all test 
cases after implementing the GCE. Here we also compare the gas fraction of our 
galaxies, both early-type and discs, with those from \citet{Kannappan:04} as a 
function of $u-r$ colour, as shown in Figure \ref{gas-frac2}. The agreement in 
the slope, scatter and zero-point of the relation is quite good for all models, 
especially when the new model of galactic chemical evolution is included.

\begin{figure}
\includegraphics[width=85mm]{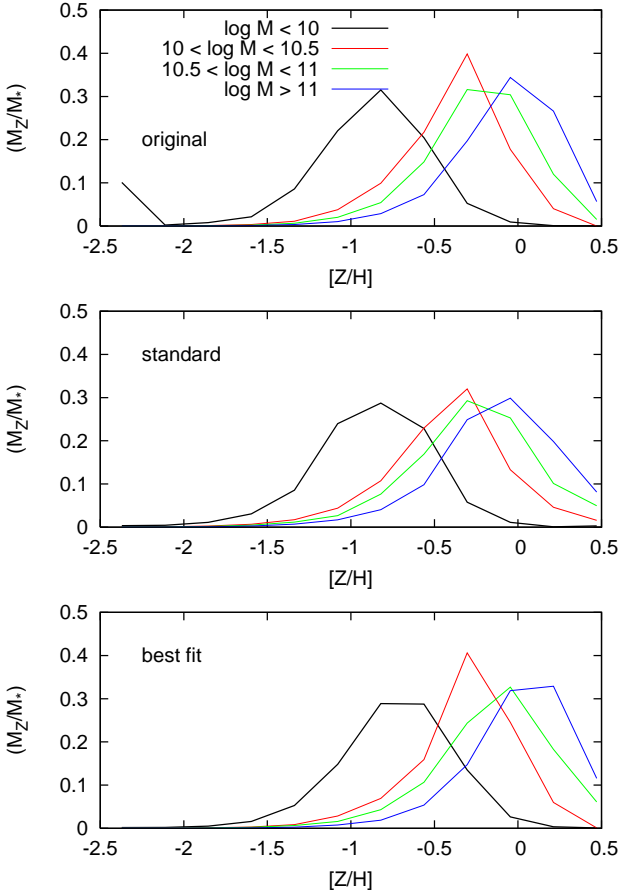}
\caption{Average metallicity distributions for model early-type galaxies in (from 
bottom to top) the SAM+GCE with the best fitting parameters ($x = 1.1$ \& 
$A=0.015$), with standard parameters ($x = 1.3$ \& $A=0.05$), and in the SAM 
without GCE, binned by stellar mass.}
\label{mdf}
\end{figure}

The first evidence for the mass-metallicity relation can be seen when looking at 
the metallicity distribution function of galaxies of different masses, which we 
show in Figure~\ref{mdf} for our three test cases. All of the models agree 
qualitatively, showing an increasing mean of the distribution as the mass range 
increases. At a given mass, however, the distributions for the best-fitting model 
are shifted to higher metallicities since galaxies in this simulation are more 
metal rich (see below).

Summarising, we have seen that including a detailed chemical evolution model in 
the SAM has a minor effect on the predicted formation histories and present-day 
properties of galaxies, and therefore does not require a re-calibration of the 
free parameters of the model. In the next subsection we will investigate the 
predicted metallicities and abundances, which are indeed affected by the GCE.

\subsection{Chemistry of Early-type Galaxies}
\label{sec:res}

The main effect of the new treatment of chemical evolution is reflected in the 
metallicity and abundance ratios of the galaxies. Most SAMs reproduce fairly well 
the mass-metallicity relation of galaxies (with effective yield treated as a free 
parameter), but to date they have been unsuccessful in fitting the slope of the 
mass--[$\alpha$/Fe] relation \citep[e.g.,][]{Nagashima05b,Pipino08}. This is the 
main challenge that we address in this study.

The galaxy sample used for comparison is that described in \citet{T00a}. This 
sample has been reanalysed using the updated stellar population synthesis method 
presented in \citet{TFD08}, which is sensitive to age, metallicity and abundance 
ratios. In the current study, we use models based on the \citet{BC03} models with 
index variations due to abundance ratios taken from \citet{Lee09}. Inferred 
stellar population parameters are tabulated in Appendix B. When making the 
comparisons, we use the stellar mass of the simulated galaxies and the inferred 
dynamical mass of the observed ones. This should not introduce a significant bias 
since the dynamical mass is a good tracer of the stellar mass within one 
effective radius for most early-type galaxies \citep{Cappellari06}. The 
abundances presented here were normalised to the solar values from 
\citet{Grevesse}. We have also computed the present epoch SNe Ia and II rates for 
our galaxies and compared them with the results of \citet{Mannucci05} and 
\citet{sullivan06}.
 
\subsubsection{Dependence on IMF slope and SN Ia fraction}

\begin{figure}
\includegraphics[width=85mm]{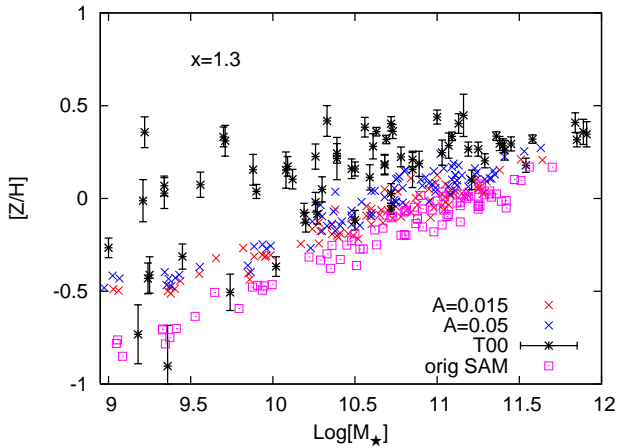}
\caption{The relationship between metallicity and stellar mass for the galaxies 
in our simulations and in the observational sample of T00. Symbols -- pink 
squares: original SAM; red crosses: SAM+GCE with $x=1.3$ and $A=0.015$; blue 
crosses: SAM+GCE with $x=1.3$ and $A=0.05$; black stars with error bars -- 
galaxies from \citet{T00a}, reanalysed as described in the text. Note the poor 
agreement of the model galaxies with the observations in all cases.}
\label{sig-Z1}
\end{figure}

\begin{figure}
\includegraphics[width=85mm]{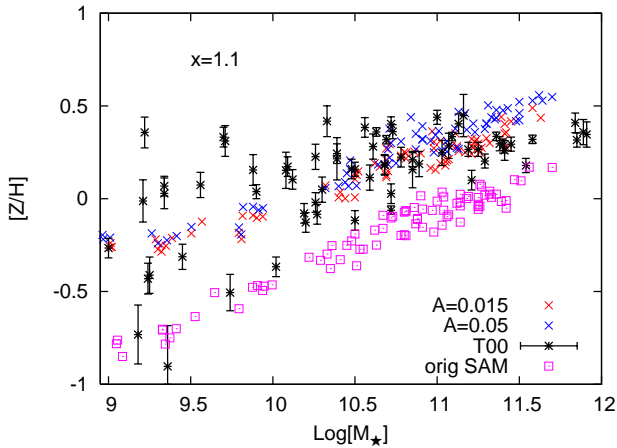}
\caption{As in Figure~\ref{sig-Z1}, except with a shallower high-mass
  slope for the IMF ($x = 1.1$) used in the SAM+GCE. Note the
  significantly better agreement of the model galaxies with the
  observations.}
\label{sig-Z2}
\end{figure}

In Figures \ref{sig-Z1} and \ref{sig-Z2} we show the relation between total 
metallicity ([Z/H]) and stellar mass ($M_{\star}$). From these figures, we see 
that in the ``old'' SAMs, galaxies tended to be too metal-poor compared to the 
observations (as seen in TS09), and implementing the detailed chemical evolution 
with the standard IMF improves the results only slightly. We explore the 
parameter space of the GCE equations to see if it is possible to improve the 
agreement with the observations. Specifically, the parameters allowed to vary 
are the fraction of binaries that give rise to SNe Ia ($A$ in Eq.~\ref{ejec}), 
the slope of the IMF above $1\,M_{\sun}$ ($x$ in Eq.~\ref{IMF}) and the upper 
mass limit of the IMF ($M_U$ in Eq.~\ref{ejec}).  However, we will only show the 
results for $M_U=40\,M_{\sun}$ for the following reason. Given that our chosen SN 
II yields (WW95) are only tabulated up to that value, we are forced to assume 
that the yields relative to the initial mass remain constant and equal to 
those of a $40\,M_{\sun}$ star for stars above this value when running 
simulations with higher upper mass limits. This proves to be an unreliable 
assumption, as reflected for example in the non-monotonic behaviour of the 
[$\alpha$/Fe] ratio as a function of the stellar mass of our galaxies, in serious 
disagreement with observations. Other properties, such as the SNe Ia rate, are 
not significantly affected by this parameter, as expected.

We also note that, even though we have varied the slope of the
  IMF, the supernova feedback efficiency remains the same because the
  prescription for the SN feedback energetics and the SN chemical
  enrichment are decoupled; the SN feedback efficiency is set manually
  as a free parameter (see Eq. \ref{reheat}), independent of the
  IMF. In this sense, the model is not entirely self-consistent.
  However, this choice makes it easier to interpret the effect of
  changing the IMF in the models. 

We now compare the predictions for the observed galaxy sample, the
original SAM, and the new SAM with GCE with a standard Chabrier IMF, a
shallower IMF ($x=1.1$) and two choices of the parameter $A$ (0.015
and 0.05). The use of a shallower IMF results in an upwards shift in
the zero-point of the mass-metallicity relation, bringing it into much
better agreement with the observations. This increase in metallicity
with a shallower IMF is expected since more massive stars are produced
and therefore the gas is enriched more efficiently. The metallicities
of the model galaxies show almost no dependence on the parameter $A$,
which is not surprising since this parameter mainly controls the ratio
of type Ia to type II supernovae, affecting abundance ratios but not
the overall metallicity (in essence the production of
$\alpha$-elements and Fe-peak elements compensate for one another).

\begin{figure}
\includegraphics[width=85mm]{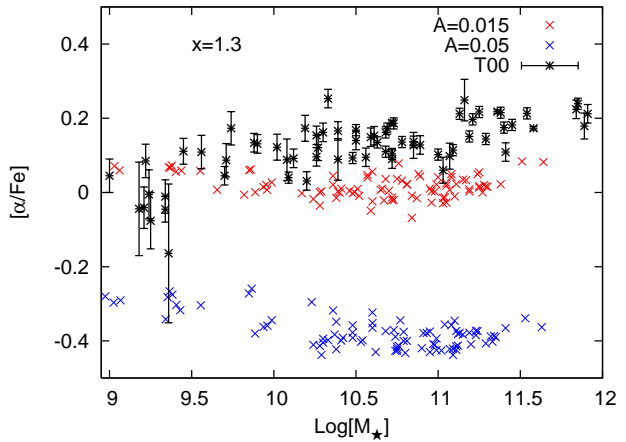}
\caption{Relation between the [$\alpha$/Fe] ratio and stellar
  mass. Symbols -- red crosses: SAM+GCE with $x=1.3$ and $A=0.015$;
  blue crosses: SAM+GCE with $x=1.3$ and $A=0.05$; black stars with
  error bars: galaxies from \citet{T00a}, reanalysed as discussed in
  the text.  Note again the poor agreement of the model galaxies with
  the observations.}
\label{sig-E1}
\end{figure}

\begin{figure}
\includegraphics[width=85mm]{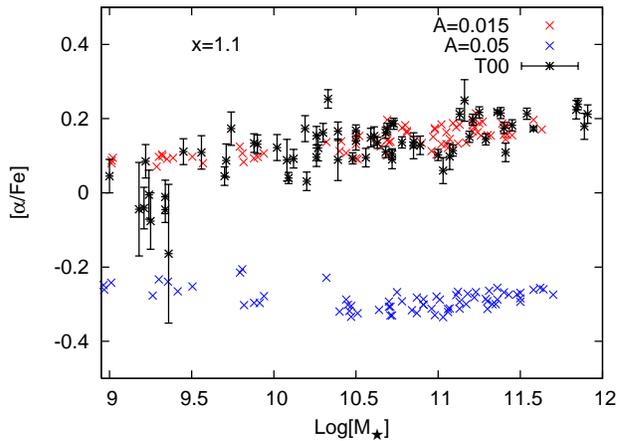}
\caption{As in Figure \ref{sig-E1}, except $x = 1.1$ for the model
  galaxies from the SAM+GCE. Note the better agreement with
  observations when using a low value of the SN Ia fraction $A$.}
\label{sig-E2}
\end{figure}

A primary advantage of our new model is that we can now calculate
abundance ratios for our galaxies, which gives us yet another property
to compare with observations and set further constraints on the
models. This is shown in Figures~\ref{sig-E1} and \ref{sig-E2}, where
we plot the [$\alpha$/Fe] ratio against stellar mass for the same
parameter choices as before. For the simulated galaxies, we consider
the abundance of $\alpha$-elements to be the composite abundance of N,
Na, Ne, Mg, Si, and S \citep[cf.][]{T00a}. If we assume a value of the
parameter $A \sim 0.05$ commonly used in the literature, the abundance
ratios are far too low. The overall values of [$\alpha$/Fe] can be
raised by decreasing the value of $A$ (i.e. producing fewer SNe Ia and
consequently less iron). However, even with this decreased $A$ and the
standard IMF, we see that the relation is too flat or even has a
slightly negative slope, and for the galaxies at the high mass end the
[$\alpha$/Fe] ratio is insufficiently enhanced.  On the other hand, a
shallower IMF, combined with a lower value of $A$, produces model
galaxies in far better agreement with observations.  The flatter
  IMF increases the slope of the relation while a low $A$ brings up
  the zero-point. Moreover, it is interesting to note that lowering
  $A$ also increases the slope (at any fixed IMF). This small steepening
  of the [$\alpha$/Fe]-mass relation is due to the metallicity
  dependence of the yields, since the higher the initial metallicity,
  the higher the [$\alpha$/Fe] in the yields. A lower fraction of SNe
  Ia (and consequently more type II) implies a slightly faster overall
  enrichment, and therefore galaxies spend more time forming stars in
  a regime of enhanced [$\alpha$/Fe] (higher metallicity yields).

In summary, we had expected that the inclusion of AGN feedback in the semi-analytic 
models might solve the problems that previous studies have encountered in trying to 
reproduce the observed trend between mass and [$\alpha$/Fe] ratio, because the quenching 
due to AGN does lead to more massive galaxies having shorter formation timescales in 
the models. However, apparently the effect of ``downsizing'' on the trend of 
[$\alpha$/Fe] is very small and a flatter IMF is required to achieve agreement between 
the models and the observations. This does not undermine the potential importance of AGN 
feedback, which appears to be a promising mechanism for solving many of the other 
problems experienced by earlier generations of models (such as the overcooling 
problem). In any case, it is encouraging that with minor variations (within the 
observational uncertainties) in the chemical evolution parameters, we can for the first 
time obtain very good agreement with the observed mass-[$\alpha$/Fe] in a semi-analytic
model.

One concern is that the comparison between our models and the
observations is not strictly rigorous since we are showing stellar
mass-weighted abundances for the models, while stellar population
studies derive abundances from line-strength indexes from integrated
spectra, which are themselves light-weighted quantities.  However,
TS09 have shown that the SSP-equivalent (absorption-line-weighted)
metallicity correlates very well with its mass-weighted and
light-weighted counterparts. In future work, nevertheless, we will
synthesise line strengths for the galaxies in our simulations and
calculate abundance ratios in the same way as is done in the
observational data.  One additional worry is the effect of limited
aperture size on the comparison, as early-type galaxies are well-known
to have significant line-strength gradients. These gradients imply however mild
\emph{metallicity} gradients but \emph{no abundance ratio} gradients
whatsoever \citep[e.g.,][TS09]{DSP93,mehlert03,SB07}. Therefore we are confident that
trends in $[\alpha/\mathrm{Fe}]$ with mass are trustworthy.  We note
here (as described in Appendix B) that the data plotted in the figures
have been constructed to appear as if the galaxies were at the
distance of the Coma cluster and observed through fibre apertures of
diameter $2\farcs7$ \citep[see][]{TFD08}. While this does not
eliminate gradient effects on the inferred metallicities, it reduces
their magnitude to an offset of roughly $-0.1$ dex (TS09).

\begin{figure}
\includegraphics[width=85mm]{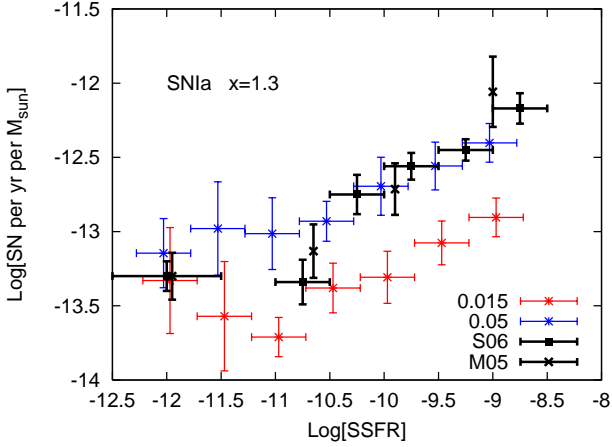}
\caption{Present day SN Ia rate as a function of specific SFR. Here we show all 
model galaxies, regardless of morphology. Red stars are SAM+GCE with $x=1.3$ and 
$A=0.015$; blue stars SAM+GCE with $x=1.3$ and $A=0.05$; black squares are 
observations from \citet{sullivan06} and black crosses from \citet{Mannucci05}. 
The conversion of galaxy type into specific SFR for the Mannucci et al. data 
points is the same as in \citet{sullivan06}.}
\label{sn-col1}
\end{figure}

\begin{figure}
\includegraphics[width=85mm]{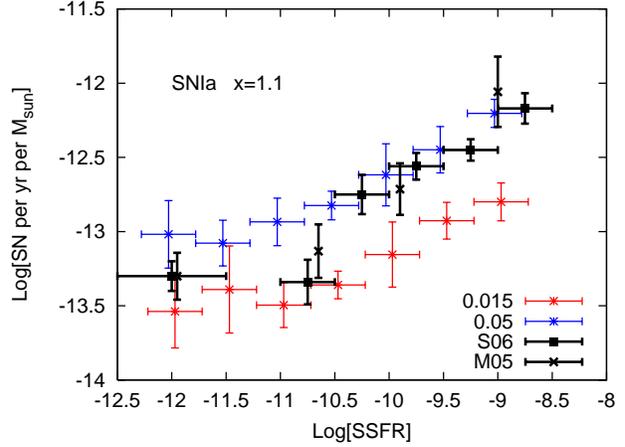}
\caption{As in Figure \ref{sn-col1}, except using the shallower IMF ($x = 1.1$) 
for the model galaxies from the SAM+GCE.}
\label{sn-col2}
\end{figure}

\begin{figure}
\includegraphics[width=85mm]{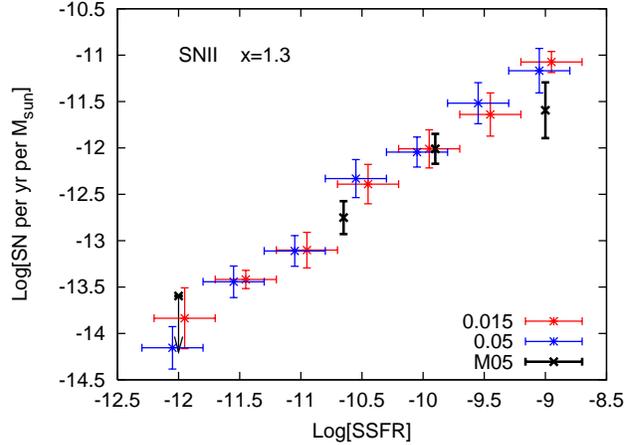}
\caption{Present-day SN II rate as a function of SSFR. Symbols are as in Figure 
\ref{sn-col1}. Results are shown for the standard IMF ($x=1.3$). We remind the 
reader that the upper-mass limit on the IMF is $40\,M_{\sun}$}
\label{snii-a}
\end{figure}

\begin{figure}
\includegraphics[width=85mm]{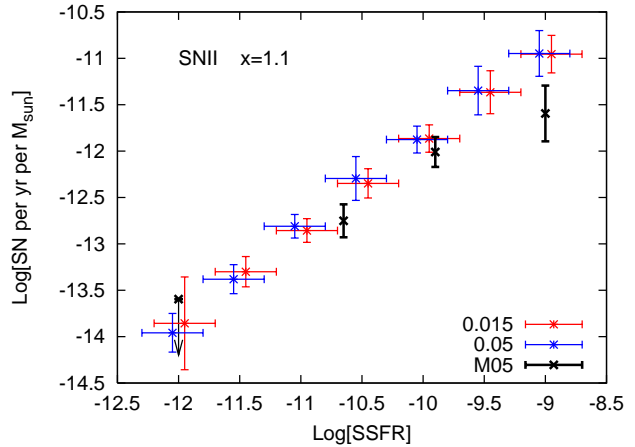}
\caption{The same as Figure \ref{snii-a}, except we use the shallower IMF slope 
($x = 1.1$) for the model galaxies from the SAM+GCE.}
\label{snii-b}
\end{figure}

Supernovae rates provide further independent constraints our models. We calculate 
the predicted supernovae rates for our model galaxies and compare them with those 
derived by \citet{Mannucci05} and \citet{sullivan06} for a large sample of 
galaxies in the nearby universe. In Figures \ref{sn-col1} and \ref{sn-col2} we 
show the present day type Ia supernova rate in units of SN events per year per 
unit stellar mass versus the specific star formation rate (SSFR, star formation 
rate per unit stellar mass), and in Figures \ref{snii-a} and \ref{snii-b} we show 
the rates for Type II SNe. Here we show all the model galaxies regardless of 
morphology, since the early-type galaxies populate only the lower SSFR side of 
these diagrams, and both comparison samples include galaxies of all morphological 
types.  The galaxies in these sample are from a mixture of field and small 
cluster environments, but so are our model galaxies.

The slope of the IMF has little effect on the predicted SNIa rates;
only the $A$ parameter has a significant influence on the
results. From Figures \ref{sn-col1} and \ref{sn-col2}, we see that no
combination of IMF slope and fraction of binaries that yield SNe Ia
can fit the observations over the whole range of SSFR.  However, it is
interesting to notice that SNe Ia rates of star forming galaxies (SSFR
$> 10^{-10.5}$) are very well matched by models with a high value of
$A$, while a low value of $A$ is a better match for passive galaxies
(SSFR $< 10^{-10.5}$).  This behaviour, which is seen regardless of
the slope of the IMF, is almost certainly due to the chosen supernova
Ia model. In the \citet{GR83} formalism which we use, the Delay Time
Distribution (DTD) of the SN Ia explosions is given by a convolution
of the distribution of secondary masses (in binary systems) and the
lifetime of the secondary star. On the other hand, from the same
observational data, \citet{MDP06} derived a DTD with two components: a
prompt peak and a later plateau, each encompassing half of the SNe
Ia. Other authors have also reached similar conclusions about the
delay-time distribution of type Ia explosions
\citep[e.g.][]{SB05,dahlen:08}. This bimodal DTD effectively enhances
the production of type Ia supernovae in star forming galaxies, exactly
where a higher fraction of SNIa are needed in our models. We therefore
expect that using a two-population DTD with a significant prompt
component will alleviate the differences between our model
predictions and the observed Type Ia rates, and we show that
  this is the case below.

The Type II SN rates, on the other hand, show very good agreement with the 
observations over the whole range of SSFR. In this case, all variations of $x$ 
and $A$ give the same qualitative results, although the models with the 
``standard'' parameter values are slightly better. Nevertheless, as we have shown 
above, not all combinations of IMF slopes and values of $A$ (SN Ia producing 
fraction of binaries) produce model galaxies that agree with observed 
metallicities and abundance ratios. It is only for those models with a shallower 
IMF ($x\sim 1.1$) and lower SNe Ia fraction among binaries ($A\sim 0.01$--0.02) 
that we can reasonably reproduce the full set of observations.

To summarise, after implementing detailed chemical evolution in the semi-analytic 
model of S08, we can now reproduce the mass--metallicity and mass--[$\alpha$/Fe] 
relations for local early-type galaxies, provided we use a slightly flatter 
Chabrier IMF and a low fraction of binaries giving rise to SN Ia. The predicted 
rates of Type Ia SNe show a strong dependence on the fraction of binaries that 
yield such an event (our parameter $A$) but not on the slope of the IMF, with the 
rates in star forming galaxies better matched by a high value of $A$, while those 
in passive galaxies are better with a low value of A. However, for a standard IMF 
and high value of $A$, the galaxies are a bit too metal-poor and, most 
importantly, the abundance ratios are extremely low, in severe disagreement with 
observations.

\subsubsection{Bimodal Delay Time Distribution for Type Ia Supernovae}

\begin{figure*}
\includegraphics[angle=270, width=170mm]{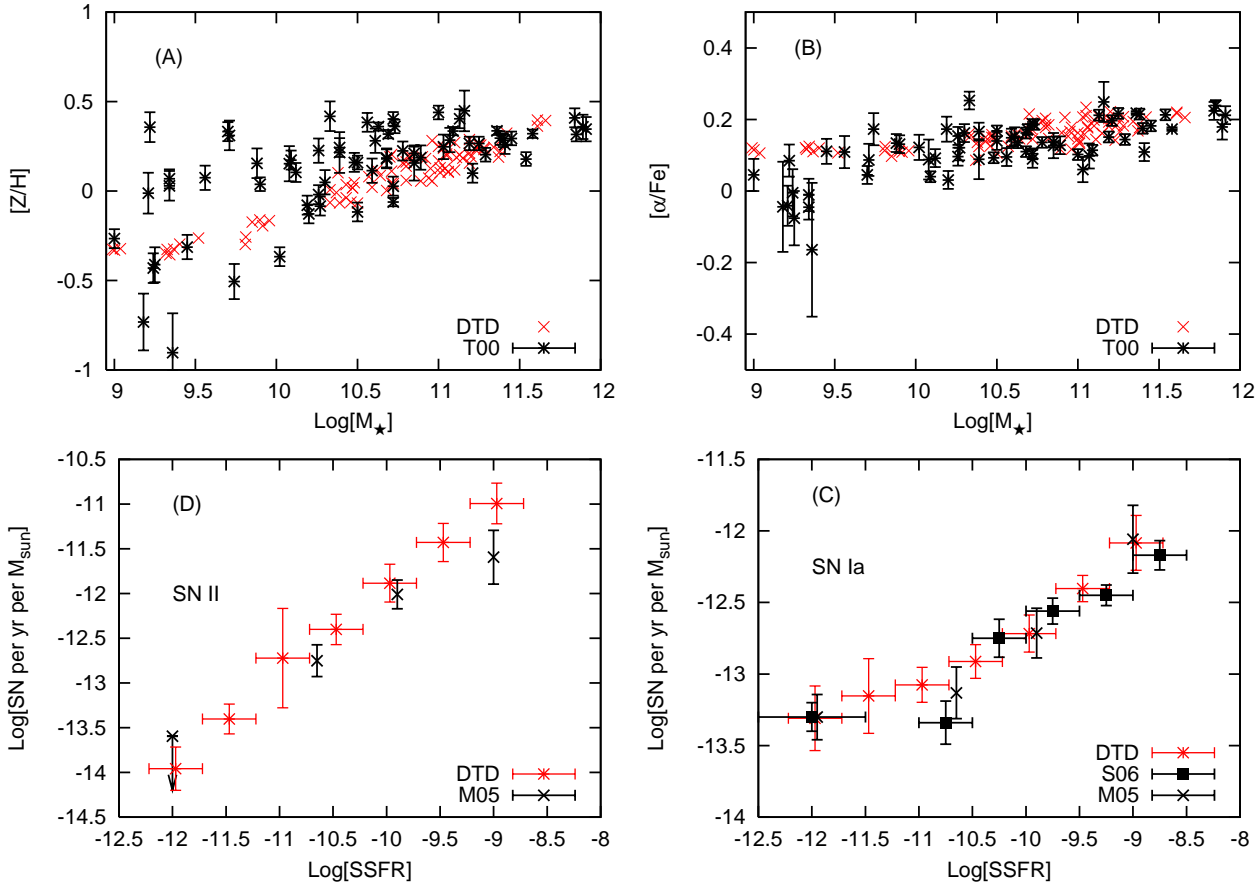}
\caption{Clockwise, starting from the top left panel: (A) [Z/H]
  vs. stellar mass; (B) [$\alpha$/Fe] vs. stellar mass; (C) Type Ia
  SNR vs. SSFR; (D) Type II SNR vs. SSFR.  Symbols -- red crosses:
  SAM+GCE with bimodal DTD for SNe Ia ($x=1.15,\,A=0.03$); T00:
  reanalysed metallicities and abundance ratios from \citet{T00a}, M05
  and S06: SN rates from \citet{Mannucci05} and \citet{sullivan06}
  respectively.}
\label{dtdall} 
\end{figure*}

In the previous section, we speculated that a bimodal
  distribution with a prompt population of SNe Ia explotions would
  give a better match for the observed supernova rates. Given the
  analytical nature of the DTD formulation \citep{greggio05}, it is
  fairly straightforward to implement and test this hypothesis. We have implemented the DTD proposed by \citet{MDP06}. Figure \ref{dtdall}
  shows the four observational constraints used to test the models
  (metallicity vs. stellar mass, [$\alpha$/Fe] vs. stellar mass, type
  Ia SNR vs. SSFR, and type II SNR vs. SSFR).  Clearly the new
  double-peaked SN Ia DTD model gives a better match to the Type Ia SN
  rates, while maintaining the good agreement in the other galactic
  properties. However, for this model, the best-fitting parameters for
  the IMF slope and the SNIa binary fraction are slightly different;
  specifically, $x=1.15$ and $A=0.03$. This value for the IMF is in
  even better agreement with some recent studies
  \citep{baldry03,wilkins08b} then our previous `best' value of
  $x=1.1$.

\begin{figure*}
\includegraphics[angle=270, width=170mm]{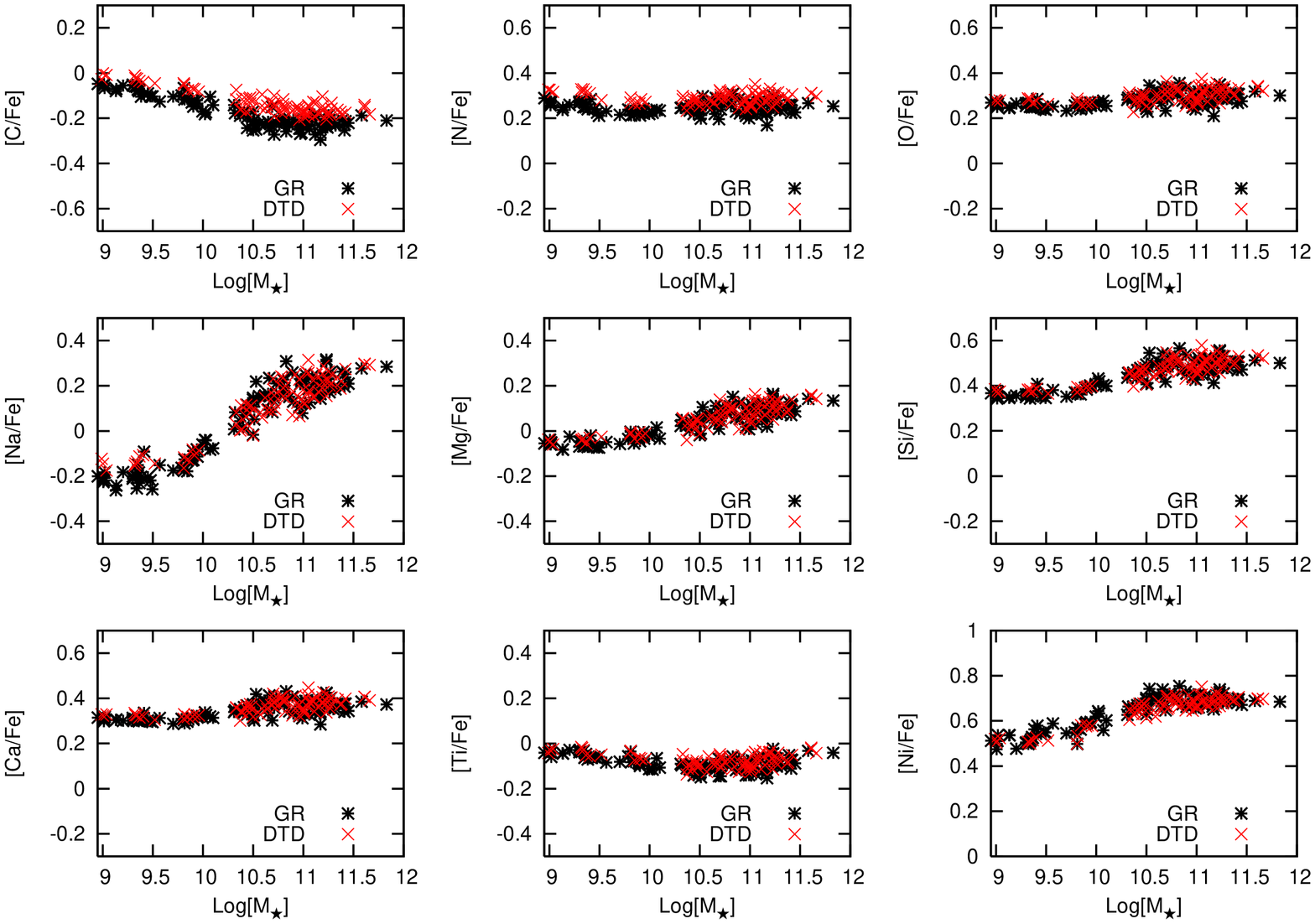}
\caption{[el/Fe] versus stellar mass for some $\alpha$-elements and
  Fe-peak elements. Only predictions from our best-fitting models are
  shown. Symbols -- black stars: Classic SN Ia \citep{GR83}; red
  crosses: DTD formulation \citep{greggio05}.  The parameters are
  ($x=1.1$, $A=0.015$) and ($x=1.15$, $A=0.03$) respectively.}
\label{abu} 
\end{figure*}

Finally, in Figure \ref{abu} we show the abundance ratios of
  some individual elements for the galaxies in our best fitting models
  using the classic SNe Ia recipe ($x=1.1$ and $A=0.015$) and the bimodal DTD
  ($x=1.15$ and $A=0.03$). With the exception of C and N, which are
  slightly higher for the latter, all the elements follow the same
  trends in both models. In particular Mg, even though it is
under-abundant with respect to other $\alpha$-elements, is the one
element that best follows the observed trend of [$\alpha$/Fe]. This
leads us to believe that the abundances derived from the stellar
population analysis may be predominantly driven by Mg, and may not
necessarily reflect of all the $\alpha$-elements.  There is also an
excess of Ni in the Fe-peak group and a decreasing trend of [C/Fe]
with increasing galactic mass which is apparently not observed
\citep{SB03,GS08}.  This should not be considered a flaw of the model,
since the abundances of individual elements are very sensitive to the
chosen yields. We expect that future line-strength observations,
interpreted with next-generation stellar population models \citep[such
  as][]{Schiavon07,Lee09}, will provide an interesting test of our
models, including the set of assumed yields (KL07 + WW95 + N97).

\subsubsection{Dependence on other model parameters}

We have also explored whether we could match the observations
investigated above by varying the galaxy formation parameters of the
SAM instead of the IMF and binary fraction parameters. For this
purpose, we ran several simulations in which we modified the star
formation efficiency ($A_K$ in Eq.~\ref{sfrec}), the SN feedback
efficiency ($\epsilon_0^{SN}$ in Eq.~\ref{reheat}) and the virial
velocity below which ejection of reheated gas from the halo into the
diffuse intergalactic medium becomes important ($V_{\mathrm{eject}}$
in Eq. \ref{expul}).  When the star formation efficiency was increased
by a constant factor, a few more massive galaxies were produced and
the slope of the mass--[$\alpha$/Fe] relation increased slightly but
the slope and zero point of the mass--metallicity relation did not
change. A higher SF efficiency implies that the cold gas is consumed
more rapidly and therefore the timescale for star formation is
shorter, which is why the mass--[$\alpha$/Fe] relation was affected.

If $A_K$ was allowed to increase with increasing galactic baryonic mass, a 
considerable number of very massive galaxies were produced and the slope of the 
mass--metallicity relation increased mildly (but not the zero point). On the 
other hand, if the factor decreased with increasing galactic mass, the trends remained the same but no galaxies above $10^{11}\,M_{\sun}$ were produced. This 
excess or lack of high mass galaxies arises because star formation in the biggest 
systems is either boosted or suppressed by this mass-dependent variation of the 
SF efficiency.

The effects of reducing the SN feedback efficiency and the $V_{\mathrm{eject}}$ 
parameter were roughly the same. In both cases the slope and zero point of the 
mass--metallicity relation increased slightly and the central galaxies of the DM 
haloes were on average more massive, but the mass--[$\alpha$/Fe] relation did not 
change and an unrealistically large number of low-mass satellite galaxies was 
also produced.  This excess of low-mass satellites is due to the fact that small 
galaxies retain their gas more efficiently when these parameters that control the 
SN feedback are decreased.

Overall, the effect of changing these other parameters is not as strong as 
flattening the IMF, and also destroys the agreement with other well-calibrated 
observations, such as the luminosity function, the metallicity distribution 
function, and the cold gas fraction. We therefore conclude that our results are 
robust to the values of these free parameters.

As a final test, we modified some of the yields. Specifically, we decreased the 
Fe yield of SNe II by half and increased the Mg yield by a factor of four. 
Reducing the Fe yield had very little effect, indicating that the bulk of the Fe 
comes from SNe Ia, as expected. Changing the Mg yield raised the zero point of 
the relations, as expected.  However, neither of these changes affected the 
slope. \citet{Pipino08} reached a similar conclusion about the yields and other 
parameters when exploring the parameter space in their models.

\section{Discussion and Conclusions}

We have implemented detailed galactic chemical evolution in a semi-analytic 
model, and use the resulting model to study the metal enrichment of early type 
galaxies in the local universe. The base SAM is that presented in \citet{s08}. We 
take into account the effects of galaxy mergers, inflow of cold gas, and SN and 
AGN driven outflows, as well as the production of metals by SN Ia, SN II and AGB 
stars. Unlike most previous SAMs we discard the instantaneous recycling 
approximation by properly accounting for the finite lifetimes of the stars and 
also make use of metallicity dependent yields. 

We run our SAM+GCE simulations in a grid of dark matter haloes ranging over 
present-day masses of $10^{11}$ to $10^{13}\,M_{\sun}$. We allow the slope of the 
IMF and the fraction of binaries that produce a SN Ia event to vary, and compare 
our results with the observed trends of metallicity and abundance ratio 
([$\alpha$/Fe]) against stellar mass of the galaxies, as well as the supernova 
rate (both type Ia and II) as a function of specific star formation rate. Only 
the models with a shallow IMF ($x=1.1$) and a low fraction of SN Ia from binaries 
($A\sim0.015$) match all four observations of early-type galaxies simultaneously. 
A slightly flatter than standard IMF is necessary in order to produce more 
massive stars, which enrich the interstellar medium more efficiently, making the 
galaxies in our simulations become more metal rich and improving the agreement 
with the data. The production of more massive stars, along with the fact that the 
star formation histories are more extended in time as the galaxy mass 
decreases, helps to achieve the correct trend of increasing [$\alpha$/Fe] with 
increasing galaxy stellar mass. However, it is also necessary to invoke a low 
fraction of SNe Ia to raise the zero-point of this relation. We also predict 
abundance patterns for a variety of elements for early-type galaxies at $z=0$ in 
our fiducial model. These predictions will be interesting to compare with future 
observations.

From studying the SNe Ia rates, we find evidence supporting a
`two-population' distribution for the type Ia explosions, since
galaxies with high specific star formation rates are better matched by
models with a high fraction of binaries that explode as SN Ia ($A$)
while those with low SSFR require a low value of $A$.   We tested
  whether the use of a more realistic (bimodal) delay-time
  distribution of type Ia supernovae would, in fact, improve the
  results. After implementing the DTD formulation for SNe Ia and using
  a bimodal distribution with a prompt peak and an extended plateau,
  we found very good agreement with the SN rates, while
  still matching the trends of [Z/H] and [$\alpha$/Fe] with stellar
  mass, although the best values for the slope of the IMF changed
  slightly and the fraction of SNeIa binaries needed to be
  doubled. Our favored model is now one with a Chabrier-like IMF with
  a slope of $x=1.15$, a SNe Ia binary fraction of $A=0.03$ (relative
  to the $3-16\,M_{\sun}$ range, $A\sim 0.0014$ relative to the full
  range of masses defining the IMF), and a bimodal
  delay-time-distribution for Type Ia SN events as proposed by
  \citet{MDP06}.
 
We have also studied the effects of varying the galaxy formation parameters in the 
SAM, but found that we were unable to reproduce the observations in this way. We 
therefore conclude that our results are robust to the values of the free 
parameters in the SAM.

This is not the first time that a GCE model has been applied within a
SAM.  \citet{Nagashima05a,Nagashima05b} have also constructed such a
model, and their `superwind' model resembles our model. They first
obtained fairly good agreement with observations of ICM abundances in
galaxy clusters, however the same models failed to reproduced the
trend of increasing [$\alpha$/Fe] with increasing galactic mass. One
of the main differences between their models and ours is in fact the
IMF. They use a Kennicutt IMF ($x=1.5$) for quiescent star formation
and a flat IMF ($x=0.0$) for stars formed in bursts. This flat IMF is
rather extreme, while the proposed modification in our models is
small, in fact within the observational uncertainties. Namely, we
require the same `shallow' IMF ($x=1.15$) for all modes of star
formation. Another model that reproduces the observed scaling of
abundance ratio with galaxy mass is that of \citet{PM04,PM06}. However
they consider a very different scenario for galaxy formation, the
monolithic collapse scenario, and allow for galaxy mergers only in the
form of a second infalling episode. More recently, \citet{Pipino08}
also coupled GCE to a SAM, but they also failed to match the
mass--[$\alpha$/Fe] relation. They claim that flattening the IMF can
not solve this problem, in contradiction with our findings. It is
worth mentioning that none of these models include AGN feedback; only
the GalICS model used by \citet{Pipino08} has some form of halo
quenching simply by shutting down the flow of cold gas onto galaxies
with masses larger than $10^{11}\,M_{\sun}$.
However, contrary to our expectations, we find that SF quenching by AGN is not a
key factor in our success at reproducing the mass--[$\alpha$/Fe] relation, even though 
the AGN feedback in our models leads to shorter formation times for the more massive 
galaxies. We find that a slight flattening of the IMF is essential to achieve agreement 
between the model and the observations. AGN feedback, nonetheless, is likely to play an 
important role in reproducing other galaxy observations, such as the stellar mass or 
luminosity function and color bimodality.

Our best-fitting IMF, nonetheless, is consistent with
  observations; the slope is within the observational uncertainty of
  \citet{Chabrier03} and agrees remarkably well with the results of
  \citet{baldry03}, who found that ultraviolet to near-infrared galaxy
  luminosity densities require an IMF with a slope of
  $1.15\pm0.2$. The same slope was found by \citet{wilkins08b} when
  trying to reconcile the redshift evolution of the observed stellar
  mass density with the cosmic SFH using a constant and universal
  IMF. The agreement, however, holds only at low redshift. In a
  forthcoming paper, Wilkins et al. (in preparation) propose an
  evolving IMF as a plausible solution. Such an IMF should be strongly
  top-heavy at high redshift (Hopkins, private
  communication). \citet{vanDokkum08} also suggests an evolving
  Chabrier-like IMF based on comparing the evolution of the $M/L$
  ratios of early-type galaxies to their colour evolution, but in this
  case the change is in the characteristic mass ($m_c$ in
  Eq. \ref{IMF}) rather than the slope, making the IMF
  ``bottom-light'' at high redshift. Such evolving IMFs could, in
  principle, work in favour of the trends of [$\alpha$/Fe] with
  stellar mass and SNR with SSFR since they produce either more SN II
  or fewer SN Ia progenitors at earlier times when massive early-type
  galaxies create most of their stars. This scenario remains to be
  tested, and moreover the issue of an evolving IMF is open to
  considerable debate given the large uncertainties on its
  constraints. On a different note, \citet{meurer09} has claimed
evidence for an IMF that depends on galactic surface brightness
(or surface density) as a plausible explanation for an observed
variation in H$_{\alpha}$/FUV flux ratio. However, they invoke
variations that are an order of magnitude larger than the deviation of
our best-fitting slope from the standard value. Finally, very
  recently, \citet{calura09} have claimed that a constant IMF cannot
  account for the trends of [Z/H] and [$\alpha$/Fe] with velocity
  dispersion in elliptical galaxies and have proposed an IMF with a
  slope depending on the SFR ($x=1.35$ for low star-forming systems
  and $x=1$ for high star-forming systems) in order to explain
  them. However, their chemical evolution is not coupled to a
  semi-analytic model, but is computed a posteriori with SFHs
  extracted from the SAM of \citet{menci08}, and therefore the flows
  of enriched gas from the galaxies into the halos and back again are
  not tracked, unlike in the our model.

In future work we will apply this model to other questions such as the
abundances of different components of spiral galaxies like the Milky
Way (e.g. the disk, bulge, and stellar halo), abundances in clusters
vs. the field, abundances in the intra-cluster gas, and to the
evolution of metals over cosmic time.

\section*{Acknowledgements}

We thank the directors of the Max-Planck-Institut f\"ur Astronomie, H.-W.~Rix, 
and the Kapteyn Astronomical Institute, J.M.~van der Hulst, and NOVA, the Lorentz 
Center and the Leids Kerkhoven-Bosscha Fonds for providing travel support and 
working space during the gestation of this paper. BKG acknowledges the support of 
the UK's Science \& Technology Facilities Council (STFC Grant ST/F002432/1) and 
the Commonwealth Cosmology Initiative. We also thank Francesca Matteucci 
for helpful advice at an early stage of this project, and the anonymous 
referee for insightful questions and commnets.

\bibliographystyle{mn2e}
\bibliography{models}

\appendix

\section{Implementation Algorithm of the GCE modelling}

\begin{table}
\caption{Structure of the grid where the star formation and abundance histories 
are stored. All quantities are in Gyr.}
\label{bining}
\begin{tabular}{@{}cc}
\hline
Age Range & Size of bins\\
\hline
0.00 -- 0.12 & 0.01 \\ 
0.12 -- 0.40 & 0.02 \\ 
0.40 -- 1.12 & 0.04 \\ 
1.12 -- 2.72 & 0.08 \\
2.72 -- 11.68 & 0.32 \\
11.68 -- & 1.28 \\
\hline
\end{tabular}

\medskip
The initial size of the bins is 0.01 Gyr because it is the maximum possible value 
of the time-step in our simulations.
\end{table}

To compute the chemical enrichment consistently as described in Section 2, it is 
imperative to keep track of the star formation history of each galaxies as well 
as the metallicity of the ISM as a function of time in these galaxies. Given the 
large number of galaxies in the simulations, it is very expensive, both in 
computing time and in physical memory, to store these quantities in a linear time 
grid with reasonable resolution. In order to overcome these limitations, we 
use an age grid where first bin of this grid represents $t=t_{\mathrm{now}}$ at 
any moment during the calculations and the size (in time) of the bins increase as 
we go to older ages. The scaling of the bins is related to the lifetimes of the 
stars, so that it increases as the lifetimes of the stars become progressively 
larger. The size of the bins only increases, but each subsequent bin is not 
necessarily larger than the previous one. In Table~\ref{bining} we show the 
structure of this binning grid. In this grid we store the amount of cold gas mass 
turned into stars, $\Delta M_{\star}$, and the metallicity of the cold gas, 
$Z_g$\footnote{The SFR at the time of interest is computed dividing $\Delta 
M_{\star}$ by the size of the bin.}. As the systems evolve this information is 
`pushed down' to older bins, directly if the subsequent bin is of the same size 
or added if the bin is of a larger size. In the later case, the cold gas 
masses are added directly whereas the the metallicity is averaged, weighted by 
the size of the bins. In essence, what we extract from this grid are 
time-averaged quantities.

It is also important how we handle this information when galaxies merge, that is, 
how we combine the star formation and metallicity histories of the merging 
galaxies. This is done on a bin-by-bin basis. At a given age bin, $\Delta 
M_{\star}$ is added directly and the metallicities are averaged weighted by the 
corresponding $\Delta M_{\star}$.  In mathematical form,
\begin{eqnarray}
\Delta M_{\star}[i]&=&\Delta M_{\star}^A[i]+\Delta M_{\star}^B[i], \\
Z_g &= &\frac{Z_g^A*\Delta M_{\star}^A[i] + Z_g^B\Delta M_{\star}^B[i]}
{\Delta M_{\star}^A[i]+\Delta M_{\star}^B[i]}, 
\end{eqnarray}
where $A$ and $B$ are the merging galaxies and $i$ the age bin. Summarising, the 
star formation rate and the cold gas metallicity used in Eq.~\ref{ejec} to 
calculate the enrichment of the ISM are average quantities.

\section{Galaxy Sample}

The sample of galaxies used to test our models was taken from \citet{T00a} and 
reanalysed with the method described in \citet{TFD08}. Specifically, we used the 
\citet{BC03} stellar population models and modified the line strengths when 
$[\alpha/\mathrm{Fe}]\neq 0$ using `response functions' that have been calculated 
for each star in each isochrone, a significant improvement over previous methods 
\citep[e.g.,][]{TB95,KMT05}. More details can be found in \citet{TFD08} and 
\citet{Lee09}. In Table \ref{data} we present the results from this new stellar 
population synthesis analysis.

We also present dynamical mass computed using
\begin{equation}
  M_{\mathrm{dyn}}=465 \sigma^2 r_e,
\end{equation}
where $\sigma$ is the central velocity dispersion in $\mathrm{km\,s^{-1}}$ (here 
we use the velocity dispersion taken through an aperture of diameter $r_e/8$) and 
$r_e$ is effective radius in parsecs \citep[see][for more details]{T00b}.

\bsp

\clearpage

\begin{table*}
\begin{minipage}{130mm}
\caption{Velocity dispersions, ages, metallicities, enhancement ratios and 
dynamical masses of the sample galaxies.}
\label{data}
\begin{tabular}{@{}lcrrrrr}
\hline
\multicolumn{1}{c}{Galaxy} & \multicolumn{1}{c}{Type} &
\multicolumn{1}{c}{Log($\sigma$)} & \multicolumn{1}{c}{log(Age)} &
\multicolumn{1}{c}{[Z/H]} & \multicolumn{1}{c}{[E/Fe]} &
\multicolumn{1}{c}{$\log(M_{\mathrm{dyn}})$} \\
\hline
ESO358-G06 & S0  &  $ 1.763 \pm 0.072 $  &  $0.807_{-0.250}^{+0.114}$  &  $
-0.431_{-0.098}^{+0.083}$  &  $ -0.005_{-0.045}^{+0.066}$  &  $  9.24 \pm 0.43 $
\\
ESO358-G25 & S0  &  $ 1.690 \pm 0.072 $  &  $0.779_{-0.295}^{+0.326}$  &  $
-0.732_{-0.144}^{+0.159}$  &  $ -0.044_{-0.096}^{+0.126}$  &  $  9.18 \pm 0.39 $
\\
ESO358-G50 & S0  &  $ 1.732 \pm 0.072 $  &  $0.707_{-0.235}^{+0.129}$  &  $
-0.412_{-0.098}^{+0.098}$  &  $ -0.076_{-0.056}^{+0.076}$  &  $  9.25 \pm 0.48 $
\\
ESO358-G59 & S0  &  $ 1.653 \pm 0.072 $  &  $0.544_{-0.068}^{+0.235}$  &  $
-0.266_{-0.114}^{+0.053}$  &  $  0.045_{-0.035}^{+0.045}$  &  $  9.00 \pm 0.45 $
\\
ESO359-G02 & S0  &  $ 1.763 \pm 0.072 $  &  $0.603_{-0.189}^{+0.159}$  &  $
-0.904_{-0.189}^{+0.220}$  &  $ -0.164_{-0.500}^{+0.187}$  &  $  9.36 \pm 0.46 $
\\
IC1963     & S0  &  $ 1.763 \pm 0.072 $  &  $0.397_{-0.068}^{+0.114}$  &  $ 
0.357_{-0.098}^{+0.083}$  &  $  0.085_{-0.035}^{+0.045}$  &  $  9.22 \pm 0.50 $
\\
IC2006     & E   &  $ 2.134 \pm 0.024 $  &  $1.126_{-0.114}^{+0.098}$  &  $ 
0.049_{-0.083}^{+0.068}$  &  $  0.162_{-0.015}^{+0.025}$  &  $ 10.30 \pm 0.33 $
\\
NGC0221    & E   &  $ 1.731 \pm 0.021 $  &  $0.975_{-0.053}^{+0.068}$  &  $
-0.398_{-0.038}^{+0.038}$  &  $ -0.026_{-0.005}^{+0.025}$  &  $  8.32 \pm 0.28 $
\\
NGC0224    & E   &  $ 2.185 \pm 0.008 $  &  $0.926_{-0.159}^{+0.129}$  &  $ 
0.281_{-0.038}^{+0.068}$  &  $  0.153_{-0.005}^{+0.025}$  &  $ 10.61 \pm 0.15 $
\\
NGC0315    & E   &  $ 2.486 \pm 0.005 $  &  $0.860_{-0.083}^{+0.098}$  &  $ 
0.316_{-0.023}^{+0.038}$  &  $  0.239_{-0.005}^{+0.015}$  &  $ 11.85 \pm 0.35 $
\\
NGC0507    & E   &  $ 2.445 \pm 0.009 $  &  $0.601_{-0.098}^{+0.129}$  &  $ 
0.359_{-0.053}^{+0.068}$  &  $  0.179_{-0.015}^{+0.035}$  &  $ 11.89 \pm 0.34 $
\\
NGC0584    & E   &  $ 2.278 \pm 0.005 $  &  $0.531_{-0.023}^{+0.053}$  &  $ 
0.318_{-0.023}^{+0.023}$  &  $  0.172_{-0.005}^{+0.015}$  &  $ 10.69 \pm 0.30 $
\\
NGC0636    & E   &  $ 2.185 \pm 0.007 $  &  $0.762_{-0.068}^{+0.098}$  &  $ 
0.160_{-0.053}^{+0.053}$  &  $  0.093_{-0.015}^{+0.015}$  &  $ 10.48 \pm 0.34 $
\\
NGC0720    & E   &  $ 2.381 \pm 0.011 $  &  $0.653_{-0.174}^{+0.326}$  &  $ 
0.448_{-0.083}^{+0.114}$  &  $  0.249_{-0.025}^{+0.056}$  &  $ 11.16 \pm 0.27 $
\\
NGC0821    & E   &  $ 2.281 \pm 0.005 $  &  $0.950_{-0.038}^{+0.068}$  &  $ 
0.210_{-0.023}^{+0.038}$  &  $  0.137_{-0.005}^{+0.015}$  &  $ 10.85 \pm 0.28 $
\\
NGC0936    & S0  &  $ 2.258 \pm 0.026 $  &  $1.110_{-0.098}^{+0.098}$  &  $
-0.084_{-0.053}^{+0.053}$  &  $  0.121_{-0.015}^{+0.035}$  &  $ 10.27 \pm 0.42 $
\\
NGC1316    & S0  &  $ 2.344 \pm 0.024 $  &  $0.510_{-0.053}^{+0.038}$  &  $ 
0.260_{-0.038}^{+0.053}$  &  $  0.109_{-0.015}^{+0.025}$  &  $ 11.41 \pm 0.22 $
\\
NGC1336    & E   &  $ 1.982 \pm 0.024 $  &  $1.073_{-0.068}^{+0.114}$  &  $
-0.367_{-0.053}^{+0.053}$  &  $  0.122_{-0.015}^{+0.035}$  &  $ 10.02 \pm 0.31 $
\\
NGC1339    & E   &  $ 2.199 \pm 0.024 $  &  $1.146_{-0.129}^{+0.098}$  &  $
-0.079_{-0.053}^{+0.053}$  &  $  0.173_{-0.015}^{+0.035}$  &  $ 10.19 \pm 0.40 $
\\
NGC1351    & E   &  $ 2.196 \pm 0.024 $  &  $1.129_{-0.083}^{+0.098}$  &  $
-0.117_{-0.038}^{+0.053}$  &  $  0.140_{-0.015}^{+0.025}$  &  $ 10.50 \pm 0.31 $
\\
NGC1373    & E   &  $ 1.875 \pm 0.024 $  &  $1.010_{-0.053}^{+0.083}$  &  $
-0.313_{-0.083}^{+0.068}$  &  $  0.111_{-0.015}^{+0.035}$  &  $  9.45 \pm 0.45 $
\\
NGC1374    & E   &  $ 2.267 \pm 0.024 $  &  $0.931_{-0.068}^{+0.083}$  &  $ 
0.114_{-0.114}^{+0.068}$  &  $  0.149_{-0.015}^{+0.025}$  &  $ 10.59 \pm 0.31 $
\\
NGC1375    & S0  &  $ 1.748 \pm 0.072 $  &  $0.323_{-0.023}^{+0.023}$  &  $ 
0.068_{-0.068}^{+0.053}$  &  $ -0.045_{-0.025}^{+0.035}$  &  $  9.34 \pm 0.40 $
\\
NGC1379    & E   &  $ 2.114 \pm 0.024 $  &  $0.913_{-0.053}^{+0.083}$  &  $
-0.020_{-0.053}^{+0.053}$  &  $  0.154_{-0.015}^{+0.025}$  &  $ 10.26 \pm 0.32 $
\\
NGC1380    & S0  &  $ 2.340 \pm 0.024 $  &  $1.008_{-0.083}^{+0.114}$  &  $ 
0.157_{-0.068}^{+0.098}$  &  $  0.127_{-0.015}^{+0.035}$  &  $ 10.85 \pm 0.28 $
\\
NGC1380A   & S0  &  $ 1.740 \pm 0.072 $  &  $0.476_{-0.083}^{+0.235}$  &  $
-0.012_{-0.098}^{+0.114}$  &  $ -0.041_{-0.035}^{+0.056}$  &  $  9.21 \pm 0.40 $
\\
NGC1381    & S0  &  $ 2.185 \pm 0.024 $  &  $0.923_{-0.038}^{+0.068}$  &  $ 
0.104_{-0.038}^{+0.053}$  &  $  0.092_{-0.005}^{+0.025}$  &  $ 10.12 \pm 0.39 $
\\
NGC1399    & E   &  $ 2.574 \pm 0.024 $  &  $1.026_{-0.053}^{+0.083}$  &  $ 
0.346_{-0.038}^{+0.068}$  &  $  0.212_{-0.015}^{+0.025}$  &  $ 11.91 \pm 0.21 $
\\
NGC1404    & E   &  $ 2.415 \pm 0.024 $  &  $0.994_{-0.053}^{+0.068}$  &  $ 
0.187_{-0.038}^{+0.068}$  &  $  0.128_{-0.005}^{+0.025}$  &  $ 10.89 \pm 0.32 $
\\
NGC1419    & E   &  $ 2.068 \pm 0.024 $  &  $1.308_{-0.220}^{+0.220}$  &  $
-0.506_{-0.189}^{+0.098}$  &  $  0.173_{-0.025}^{+0.045}$  &  $  9.74 \pm 0.45 $
\\
NGC1427    & E   &  $ 2.243 \pm 0.024 $  &  $0.981_{-0.023}^{+0.038}$  &  $
-0.062_{-0.023}^{+0.023}$  &  $  0.104_{-0.005}^{+0.015}$  &  $ 10.72 \pm 0.29 $
\\
NGC1453    & E   &  $ 2.475 \pm 0.005 $  &  $1.008_{-0.053}^{+0.053}$  &  $ 
0.294_{-0.023}^{+0.038}$  &  $  0.182_{-0.005}^{+0.015}$  &  $ 11.45 \pm 0.39 $
\\
NGC1461    & S0  &  $ 2.294 \pm 0.011 $  &  $0.864_{-0.098}^{+0.098}$  &  $ 
0.155_{-0.053}^{+0.083}$  &  $  0.134_{-0.015}^{+0.025}$  &  $  9.88 \pm 0.67 $
\\
NGC1600    & E   &  $ 2.508 \pm 0.009 $  &  $0.822_{-0.114}^{+0.189}$  &  $ 
0.409_{-0.023}^{+0.053}$  &  $  0.224_{-0.015}^{+0.025}$  &  $ 11.84 \pm 0.36 $
\\
NGC1700    & E   &  $ 2.356 \pm 0.005 $  &  $0.519_{-0.008}^{+0.038}$  &  $ 
0.335_{-0.023}^{+0.023}$  &  $  0.114_{-0.005}^{+0.015}$  &  $ 11.09 \pm 0.32 $
\\
NGC2300    & E   &  $ 2.418 \pm 0.006 $  &  $0.719_{-0.144}^{+0.053}$  &  $ 
0.404_{-0.023}^{+0.053}$  &  $  0.212_{-0.005}^{+0.015}$  &  $ 11.13 \pm 0.38 $
\\
NGC2560    & S0  &  $ 2.303 \pm 0.006 $  &  $0.810_{-0.174}^{+0.159}$  &  $ 
0.226_{-0.053}^{+0.068}$  &  $  0.096_{-0.015}^{+0.025}$  &  $ 10.26 \pm 0.94 $
\\
NGC2778    & E   &  $ 2.198 \pm 0.007 $  &  $0.897_{-0.174}^{+0.098}$  &  $ 
0.243_{-0.053}^{+0.053}$  &  $  0.166_{-0.005}^{+0.025}$  &  $ 10.39 \pm 0.35 $
\\
NGC3115    & S0  &  $ 2.378 \pm 0.053 $  &  $0.574_{-0.068}^{+0.159}$  &  $ 
0.384_{-0.053}^{+0.053}$  &  $  0.095_{-0.025}^{+0.025}$  &  $ 10.56 \pm 0.32 $
\\
NGC3377    & E   &  $ 1.981 \pm 0.009 $  &  $0.653_{-0.098}^{+0.068}$  &  $ 
0.038_{-0.038}^{+0.038}$  &  $  0.131_{-0.005}^{+0.025}$  &  $  9.90 \pm 0.28 $
\\
NGC3379    & E   &  $ 2.291 \pm 0.005 $  &  $0.969_{-0.023}^{+0.038}$  &  $ 
0.159_{-0.023}^{+0.038}$  &  $  0.168_{-0.005}^{+0.015}$  &  $ 10.50 \pm 0.28 $
\\
NGC3384    & S0  &  $ 2.140 \pm 0.031 $  &  $0.522_{-0.053}^{+0.083}$  &  $ 
0.332_{-0.038}^{+0.053}$  &  $  0.045_{-0.015}^{+0.025}$  &  $  9.70 \pm 0.44 $
\\
NGC3412    & S0  &  $ 2.013 \pm 0.034 $  &  $0.632_{-0.159}^{+0.098}$  &  $ 
0.029_{-0.083}^{+0.083}$  &  $ -0.011_{-0.035}^{+0.045}$  &  $  9.34 \pm 0.49 $
\\
NGC3585    & S0  &  $ 2.307 \pm 0.122 $  &  $0.710_{-0.159}^{+0.129}$  &  $ 
0.247_{-0.053}^{+0.068}$  &  $  0.060_{-0.015}^{+0.035}$  &  $ 11.03 \pm 0.35 $
\\
NGC3607    & S0  &  $ 2.332 \pm 0.101 $  &  $0.720_{-0.174}^{+0.189}$  &  $ 
0.285_{-0.068}^{+0.068}$  &  $  0.098_{-0.025}^{+0.035}$  &  $ 11.07 \pm 0.33 $
\\
NGC3608    & E   &  $ 2.259 \pm 0.006 $  &  $0.930_{-0.083}^{+0.083}$  &  $ 
0.224_{-0.038}^{+0.053}$  &  $  0.136_{-0.005}^{+0.015}$  &  $ 10.78 \pm 0.28 $
\\
NGC3818    & E   &  $ 2.246 \pm 0.007 $  &  $0.763_{-0.068}^{+0.144}$  &  $ 
0.361_{-0.023}^{+0.038}$  &  $  0.186_{-0.005}^{+0.015}$  &  $ 10.73 \pm 0.35 $
\\
NGC3941    & S0  &  $ 2.117 \pm 0.036 $  &  $0.422_{-0.098}^{+0.083}$  &  $ 
0.311_{-0.098}^{+0.083}$  &  $  0.087_{-0.035}^{+0.045}$  &  $  9.71 \pm 0.42 $
\\
NGC4026    & S0  &  $ 2.258 \pm 0.029 $  &  $0.542_{-0.083}^{+0.159}$  &  $ 
0.153_{-0.083}^{+0.098}$  &  $  0.088_{-0.035}^{+0.056}$  &  $ 10.08 \pm 0.41 $
\\
NGC4036    & S0  &  $ 2.220 \pm 0.039 $  &  $0.574_{-0.129}^{+0.114}$  &  $ 
0.418_{-0.038}^{+0.083}$  &  $  0.253_{-0.015}^{+0.025}$  &  $ 10.33 \pm 0.45 $
\\
NGC4111    & S0  &  $ 2.127 \pm 0.019 $  &  $0.464_{-0.008}^{+0.023}$  &  $ 
0.074_{-0.083}^{+0.068}$  &  $  0.109_{-0.035}^{+0.045}$  &  $  9.56 \pm 0.56 $
\\
NGC4251    & S0  &  $ 2.072 \pm 0.074 $  &  $0.514_{-0.038}^{+0.038}$  &  $ 
0.173_{-0.038}^{+0.053}$  &  $  0.040_{-0.015}^{+0.015}$  &  $ 10.09 \pm 0.38 $
\\
NGC4261    & E   &  $ 2.479 \pm 0.005 $  &  $1.136_{-0.053}^{+0.053}$  &  $ 
0.284_{-0.023}^{+0.038}$  &  $  0.175_{-0.005}^{+0.015}$  &  $ 11.40 \pm 0.28 $
\\
NGC4697    & E   &  $ 2.211 \pm 0.009 $  &  $0.924_{-0.053}^{+0.068}$  &  $ 
0.027_{-0.038}^{+0.053}$  &  $  0.090_{-0.005}^{+0.025}$  &  $ 10.72 \pm 0.23 $
\\
NGC5638    & E   &  $ 2.186 \pm 0.006 $  &  $0.946_{-0.038}^{+0.068}$  &  $ 
0.185_{-0.023}^{+0.053}$  &  $  0.162_{-0.005}^{+0.015}$  &  $ 10.68 \pm 0.29 $
\\
NGC5812    & E   &  $ 2.300 \pm 0.005 $  &  $0.602_{-0.114}^{+0.235}$  &  $ 
0.403_{-0.098}^{+0.038}$  &  $  0.188_{-0.025}^{+0.005}$  &  $ 10.72 \pm 0.33 $
\\
NGC5813    & E   &  $ 2.327 \pm 0.007 $  &  $1.127_{-0.083}^{+0.098}$  &  $ 
0.100_{-0.038}^{+0.053}$  &  $  0.197_{-0.005}^{+0.015}$  &  $ 11.21 \pm 0.26 $
\\
\hline
\end{tabular} 
\end{minipage}
\end{table*}

\setcounter{table}{0}
\begin{table*}
\begin{minipage}{130mm}
\caption{Continued.}
\begin{tabular}{@{}lcrrrrr}
\hline
\multicolumn{1}{c}{Galaxy} & \multicolumn{1}{c}{Type} &
\multicolumn{1}{c}{Log($\sigma$)} & \multicolumn{1}{c}{log(Age)} &
\multicolumn{1}{c}{[Z/H]} & \multicolumn{1}{c}{[E/Fe]} &
\multicolumn{1}{c}{$\log(M_{\mathrm{dyn}})$} \\
\hline
NGC5831    & E   &  $ 2.206 \pm 0.004 $  &  $0.560_{-0.023}^{+0.023}$  &  $ 
0.360_{-0.008}^{+0.023}$  &  $  0.135_{-0.005}^{+0.015}$  &  $ 10.63 \pm 0.31 $
\\
NGC5846    & E   &  $ 2.354 \pm 0.005 $  &  $0.980_{-0.038}^{+0.068}$  &  $ 
0.296_{-0.023}^{+0.038}$  &  $  0.215_{-0.005}^{+0.015}$  &  $ 11.38 \pm 0.23 $
\\
NGC5866    & S0  &  $ 2.143 \pm 0.116 $  &  $0.413_{-0.098}^{+0.098}$  &  $ 
0.215_{-0.114}^{+0.114}$  &  $  0.089_{-0.045}^{+0.056}$  &  $ 10.39 \pm 0.36 $
\\
NGC6127    & E   &  $ 2.393 \pm 0.007 $  &  $0.981_{-0.038}^{+0.053}$  &  $ 
0.266_{-0.023}^{+0.038}$  &  $  0.217_{-0.005}^{+0.015}$  &  $ 11.25 \pm 0.41 $
\\
NGC6702    & E   &  $ 2.243 \pm 0.005 $  &  $0.325_{-0.008}^{+0.038}$  &  $ 
0.439_{-0.023}^{+0.038}$  &  $  0.102_{-0.015}^{+0.015}$  &  $ 11.00 \pm 0.30 $
\\
NGC6703    & E   &  $ 2.258 \pm 0.006 $  &  $0.790_{-0.068}^{+0.083}$  &  $ 
0.180_{-0.023}^{+0.053}$  &  $  0.110_{-0.005}^{+0.015}$  &  $ 10.68 \pm 0.32 $
\\
NGC7052    & E   &  $ 2.464 \pm 0.006 $  &  $1.218_{-0.068}^{+0.068}$  &  $ 
0.179_{-0.038}^{+0.038}$  &  $  0.213_{-0.005}^{+0.015}$  &  $ 11.54 \pm 0.38 $
\\
NGC7454    & E   &  $ 2.028 \pm 0.011 $  &  $0.692_{-0.144}^{+0.053}$  &  $
-0.128_{-0.038}^{+0.053}$  &  $  0.031_{-0.015}^{+0.025}$  &  $ 10.20 \pm 0.31 $
\\
NGC7562    & E   &  $ 2.391 \pm 0.003 $  &  $0.924_{-0.023}^{+0.023}$  &  $ 
0.203_{-0.008}^{+0.038}$  &  $  0.144_{-0.005}^{+0.015}$  &  $ 11.29 \pm 0.35 $
\\
NGC7619    & E   &  $ 2.498 \pm 0.004 $  &  $1.051_{-0.023}^{+0.053}$  &  $ 
0.320_{-0.023}^{+0.023}$  &  $  0.173_{-0.005}^{+0.005}$  &  $ 11.58 \pm 0.30 $
\\
NGC7626    & E   &  $ 2.414 \pm 0.004 $  &  $0.963_{-0.023}^{+0.038}$  &  $ 
0.336_{-0.008}^{+0.023}$  &  $  0.218_{-0.005}^{+0.005}$  &  $ 11.36 \pm 0.37 $
\\
NGC7785    & E   &  $ 2.378 \pm 0.005 $  &  $0.911_{-0.038}^{+0.068}$  &  $ 
0.266_{-0.008}^{+0.038}$  &  $  0.151_{-0.005}^{+0.015}$  &  $ 11.19 \pm 0.39 $
\\
\hline
\end{tabular} 
\end{minipage}
\end{table*}

\label{lastpage}

\clearpage
\end{document}